\documentclass[journal]{IEEEtran}
\usepackage{algorithm}
\usepackage{algorithmicx}
\usepackage{algpseudocode}
\usepackage{amsmath}
\usepackage{comment}
\usepackage{amsthm}
\usepackage{pifont}
\usepackage{makecell}
\usepackage{arydshln}
\usepackage{booktabs}
\usepackage{multirow}
\usepackage[utf8]{inputenc}
\usepackage{tikz}
\usepackage{notoccite}
\usepackage{tabularx}
\usepackage{graphicx}
\usepackage{textcomp}
\usepackage{url}
\usepackage{hyperref}
\usepackage[most]{tcolorbox}
\usepackage{xcolor}
\usepackage{listings}
\lstdefinelanguage{solidity}{
  morekeywords={pragma, solidity, contract, library, interface, function,
    modifier, constructor, public, private, internal, external, view, pure,
    payable, returns, return, memory, storage, calldata, if, else, for, while,
    do, break, continue, require, assert, revert, emit, new, this, super,
    mapping, address, uint, uint256, uint160, uint128, uint8, int, int256,
    bool, string, bytes, bytes32, bytes4, true, false, constant, immutable,
    virtual, override, abstract, using, import, is, as, unchecked, try, catch,
    event, struct, enum},
  sensitive=true,
  morecomment=[l]{//},
  morecomment=[s]{/*}{*/},
  morestring=[b]",
  keywordstyle=\color{blue}\bfseries,
  commentstyle=\color{green!50!black}\itshape,
  stringstyle=\color{red},
}

\AtBeginDocument{%
  }
    
\begin{document}

\title{DeFiScreener: Efficient DeFi Attack Pre-screening in Smart Contracts via Historical Case Matching}

\author{Rui Cao, Shaojing Fan, Zhimei Sui, Liming Fang, Ziqi Yang, Yingying Jiao, Zhenguang Liu

\thanks{Rui Cao and Liming Fang are with the School of Computer Science and Technology, Nanjing University of Aeronautics and Astronautics, China. (e-mail: caorui@nuaa.edu.cn, fangliming@nuaa.edu.cn)}
\thanks{Shaojing Fan is with the Department of Electrical and Computer Engineering (ECE), National University of Singapore, Singapore. (e-mail: dcsfs@nus.edu.sg)}
\thanks{Zhimei Sui is with the School of Computer Engineering and Science, Shanghai University, China. (e-mail: zmsui@shu.edu.cn)}
\thanks{Ziqi Yang is with the School of Computer Science and Technology, Zhejiang University, China. (e-mail: yangziqi@zju.edu.cn)}
\thanks{Yingying Jiao is with the Zhejiang University of Technology, China. (e-mail: yingyingjiao21@gmail.com)}
\thanks{Zhenguang Liu is with the Zhejiang University (The State Key Laboratory of Blockchain and Data Security, Zhejiang University), and with Hangzhou High-Tech Zone (Binjiang) Institute of Blockchain and Data Security, China. (e-mail: liuzhenguang2008@gmail.com)}
\thanks{Corresponding Authors: Liming Fang, Zhenguang Liu and Yingying Jiao.}
}

\maketitle

\begin{abstract}
Blockchain and its killer applications, particularly decentralized finance (DeFi), are gaining widespread adoption, with over 5,200 DeFi projects deployed on mainstream blockchains as of January 2026.
At the same time, security risks in DeFi are becoming increasingly serious. However, existing DeFi detection tools usually cover only specific attack types, exhibiting severely limited detection coverage.

In this paper, we argue that an effective way to address this gap is to pre-screen vulnerable instances from large volumes of smart contract functions and call sequences. This is motivated by a key phenomenon we term \emph{``perilous temporal asymmetry''}: newly discovered attack patterns are rapidly reused on similar contracts before protections are deployed. This lag allows adversaries to reuse exploits as blueprints to target similar contracts at scale. Consequently, effective defenses must \emph{race against time}, making pre-screening a practical and proactive solution.

Inspired by this, we propose DeFiScreener, the first automated pre-screening framework for DeFi attacks that uses historical exploit cases to identify potentially vulnerable functions and call sequences. Given the full source code of a target project, DeFiScreener builds Function Call Trees (FCTs) and generates semantic embeddings for each function using a large language model (LLM), allowing both program structure and function intent to be analyzed together. It then applies a dual-level screening process. At the function level, function embeddings are matched against an Attack Pattern Library of historically exploited functions. At the sequence level, the proposed Attack Pattern Oriented Monte Carlo Tree Search (APO-MCTS) efficiently explores the \textsf{FCT}s and screens vulnerable call sequences. The identified candidates are ultimately passed to an LLM for further interpretive and security analysis.

We empirically evaluate the DeFiScreener over datasets comprising 207 real-world DeFi attack incidents, which account for approximately \$2.97 billion in aggregate losses. Experimental results demonstrate that DeFiScreener achieves a remarkable 98.55\% recall and 84.30\% precision in attack pre-screening. Interestingly, it effectively recovers 14 of 17 real-world DeFi attacks in zero-day scenario and efficiently screens 88 DeFi attacks overlooked by state-of-the-art tools.
\end{abstract}

\begin{IEEEkeywords}
Decentralized finance, smart contract, Monte Carlo tree search, DeFi attack pre-screening
\end{IEEEkeywords}

\section{Introduction}
\label{sec:intro}
\IEEEPARstart{D}{ecentralized} finance \cite{werner2022sok}, commonly referred to as DeFi, has been rapidly growing in permissionless blockchains and has recently gained significant popularity. Success stories of decentralized finance projects are reported almost daily. As of January 2026, over 5,200 DeFi projects are deployed on mainstream blockchains, with the total value locked (TVL) consistently exceeding \$120 billion \cite{defillama}. 


However, this rapidly evolving ecosystem is overshadowed by an escalating landscape of security threats. The complexity and composability of DeFi protocols have given rise to a plethora of sophisticated attacks, ranging from \textit{reentrancy} \cite{zheng2023turn, zhang2022reentrancy} to \textit{price manipulation} \cite{wu2023defiranger, xie2024defort, kong2023defitainter, zhong2025detecting} and \textit{rug pull exploits} \cite{lin2024crpwarner, zhou2024stop}. For smart contract auditors, conducting a comprehensive security audit remains an exorbitant and labor-intensive endeavor. Such an exhaustive manual review requires not only profound expertise in formal methods and program analysis but also an intimate understanding of intricate financial logics.

\textbf{Existing works and limitations.} Researchers have developed a diverse range of automated security tools. However, existing solutions predominantly focus on low-level programming flaws, rather than complex protocol-level exploits. \cite{chaliasos2024smart}. Conventional general-purpose static analysis and symbolic execution frameworks, such as Mythril \cite{mueller2018smashing}, Slither \cite{feist2019slither} and SmartCheck \cite{tikhomirov2018smartcheck}, are primarily engineered to identify classic vulnerabilities including reentrancy, timestamp dependency, and integer overflows. While these tools are effective at the instruction level, they often lack the semantic depth required to capture the high-level logic and intricate cross-contract interactions in DeFi protocols. Consequently, they remain largely inadequate for detecting sophisticated DeFi-specific threats—such as price oracle manipulation or flash loan exploits—which typically manifest as flaws in economic and protocol design rather than simple coding errors.

On the other hand, specialized DeFi attack detection methodologies, while more targeted, face significant challenges regarding coverage and maintainability. Current state-of-the-art methods often prioritize monolithic attack vectors; for instance, DeFort \cite{xie2024defort} and DeFiRanger \cite{wu2023defiranger} are tailored for price manipulation, POMABuster\cite{xi2024pomabuster} and VeriOracle \cite{mo2023toward} focuses on price oracle manipulation attacks, while CRPWarner \cite{lin2024crpwarner} and Tokeer \cite{zhou2024stop} are tailored for rug pull exploits. Achieving comprehensive coverage across the diverse DeFi landscape through such fragmented approaches is prohibitively expensive and resource-intensive for auditors. Furthermore, the rapid evolution of DeFi protocols and the continuous emergence of novel attack patterns impose a heavy maintenance burden, underscoring the pressing need for a more flexible and automated detection paradigm.

\textbf{Motivation.} Exhaustively applying single-attack detection tools is inefficient and does not scale with the growing complexity of DeFi systems, motivating the need for a pre-screening methodology. Such an approach can efficiently triage potential attack categories within smart contracts, guiding specialized detectors toward focused and high-impact analysis. This need is further underscored by a critical phenomenon we term \emph{``perilous temporal asymmetry''}: a gap between the public disclosure of a new DeFi attack and the widespread deployment of its corresponding patch. During this window, newly revealed exploits act as blueprints, enabling adversaries to rapidly scan for and target similarly structured contracts \cite{liu2025know}. In this ``race against time'', attackers can systematically exploit unpatched or replicated contracts before effective defenses are in place. This window can be measured in hours, not months: a forked BNB Chain protocol was struck just seven hours after an almost identical exploit drained its sibling contract (Section \ref{sec:motivation}).

\textbf{Our method.} Motivated by the phenomenon of recurrent attack patterns, we introduce DeFiScreener, the first automated DeFi attack pre-screening framework that leverages historical attack cases to screen potentially vulnerable functions and call sequences. Given the full source code of a target DeFi project, DeFiScreener first constructs  Function Call Trees (FCTs) from static function call relationships and extracts per-function semantic embeddings via an LLM to jointly model the protocol's structural topology and semantic intent. It then applies a dual-level screening mechanism: at the function level, we match each function embeddings against an attack pattern library populated with semantic embeddings from functions implicated in historical exploits to identify vulnerable functions; at the sequence level, we propose Attack Pattern-Oriented Monte Carlo Tree Search (APO-MCTS), which uses the historical attack patterns as prior knowledge to guide \textsf{FCT} traversal and enumerate vulnerable function call sequences. Finally, identified vulnerable functions and sequences, together with their corresponding historical attack cases, are submitted to an LLM for interpretive analysis, producing actionable, human-readable candidate vulnerability explanations for smart contract auditors.




In summary, our key contributions are:
\begin{itemize}
\item 
We propose DeFiScreener, a pre-screening framework for DeFi attack screening grounded in historical attack case matching. To the best of our knowledge, DeFiScreener is the first general pre-screening framework that discovers DeFi attacks overlooked by diverse downstream detection tools. Because this matching is pattern-agnostic, the same mechanism screens for any attack type with a documented historical precedent, rather than the one or two classes each specialized detector targets.


\item We propose APO-MCTS, an attack pattern oriented Monte Carlo Tree Search that enables pinpoint extraction of vulnerable function call sequences within seconds. 

\item We construct a high-quality dataset encompassing 207 real-world DeFi attack incidents, meticulously annotating the vulnerable functions and call sequences. Through a large-scale empirical evaluation involving incidents with an aggregate loss of approximately \$2.97 billion, we demonstrate that DeFiScreener achieves a remarkable 98.55\% recall and 84.30\% precision in attack pre-screening. Furthermore, DeFiScreener effectively recovers 14 of 17 real-world DeFi attacks
in zero-day scenario and efficiently screens 88 attacks overlooked by state-of-the-art tools.
\end{itemize}

\section{Background}
Our background section encompasses an overview of DeFi and smart contracts, as well as an introduction to DeFi attacks.

\subsection{DeFi and Smart Contracts}
DeFi represents a nascent financial infrastructure constructed upon blockchain technology, aimed at delivering transparent and permissionless financial services—including automated market making (AMM), collateralized lending, and asset management—without the intervention of traditional centralized intermediaries \cite{werner2022sok}.

The operational core of the DeFi ecosystem is comprised of Smart Contracts. These are autonomous, self-executing programs typically written in high-level languages (e.g., Solidity) and deployed on distributed ledgers like Ethereum. Once deployed, the execution of a smart contract is governed by the consensus protocol of the underlying blockchain, ensuring state transition determinism and transaction atomicity \cite{Ethereum-yellow-paper}.



\subsection{DeFi Attacks}
\label{sec:defi-attacks}
We meticulously analyze 573 DeFi attack incidents disclosed on SlowMist~\cite{slowmist} and DeFiHackLabs~\cite{DeFiHackLabs}. To organize this landscape, we group DeFi exploits into two broad categories according to their underlying vulnerability vector. These are organizing umbrella classes rather than an exhaustive taxonomy, and each subsumes many finer-grained attack types.

\subsubsection{Business Logic Vulnerabilities}
This category encompasses flaws within the smart contract’s functional implementation, such as access control, arbitrary call, and insufficient validation. These vulnerabilities typically arise when the contract fails to strictly enforce identity verification or when public-facing functions allow malicious callers to manipulate the contract’s internal state \cite{wen2024foray}.

\subsubsection{Price and Oracle Manipulations}
Such attacks exploit the financial logic and market mechanisms of DeFi protocols. Attackers often leverage large-scale liquidity provided by flash loans to manipulate asset prices on decentralized exchanges (DEXs) \cite{wu2023defiranger} or exploit dependencies on centralized price oracles \cite{xi2024pomabuster}. Such manipulations create temporary arbitrage opportunities or cause collateral insolvency, leading to significant fund drainage.

These two broad categories subsume the 26 specific attack types cataloged in Table~\ref{tab:RQ1-1}, including reentrancy, rug pull, flash-loan, and oracle-manipulation attacks. We use these categories only to structure the attack pattern library. DeFiScreener itself is pattern-agnostic: it screens any attack type for which a historical case exists, and is not limited to two classes.


\section{Motivation}
\label{sec:motivation}

Our design is motivated by the observation that many DeFi attacks are not isolated incidents, but recur across similar contracts. Indeed, the rapid evolution of the DeFi ecosystem is largely driven by an open-source development model, where developers frequently fork established protocols to accelerate deployment. While this practice fosters innovation, it also introduces systemic risk: critical vulnerabilities can be inherited and propagated across numerous derivative projects. As a result, many emerging exploits are not entirely novel, but re-implementations of previously known flaws. In this section, we present two case studies to illustrate this phenomenon, followed by a discussion of the insights that inform our design.

\subsection{Case Study I: Same-Day Exploit Replication}
A compelling illustration of this recurrence is the pair of attacks that struck the BNB Chain projects \textit{FFIST} and \textit{Utopia} on the same date—July 20, 2023~\cite{FFIST, Utopia}. Both protocols implement a token distribution mechanism via the \texttt{\_airdrop} function, which is invoked on every transfer. The function iterates a fixed number of times, deriving a series of pseudo-random airdrop addresses from on-chain parameters and forcibly overwriting each derived address's balance to exactly 1 wei.
 
\begin{figure}[htbp]
  \centering
  \begin{lstlisting}
// Shared _airdrop function (FFIST / Utopia)
function _airdrop(address from, address to, uint256 tAmount) private {
    uint256 num = 4; // FFIST: num = 1
    uint256 seed = (uint160(lastAirdropAddress) | block.number) ^ (uint160(from) ^ uint160(to));
    uint256 airdropAmount = 1;
    address airdropAddress;
    for (uint256 i; i < num;) {
        airdropAddress = address(uint160(seed | tAmount));
        _balances[airdropAddress] = airdropAmount;        // forced overwrite
        emit Transfer(airdropAddress, airdropAddress, airdropAmount);
        unchecked {
            ++i;
            seed = seed >> 1;
        }
    }
    lastAirdropAddress = airdropAddress;
}
  \end{lstlisting}
  \caption{Identical \texttt{\_airdrop} function shared by FFIST and Utopia, containing the exploitable balance-overwrite logic.}
  \label{fig:airdrop}
\end{figure}
 
As shown in Fig.~\ref{fig:airdrop}, the core vulnerability resides in the unconditional assignment \texttt{\_balances[airdropAddress] = airdropAmount} (line~9), which overwrites any target address's balance with 1 wei. Because the seed depends only on publicly observable or attacker-controlled parameters, an adversary can compute off-chain the exact \texttt{tAmount} for which a derived address coincides with the liquidity pair. This overwrites the pair's token reserve to 1 wei. The attack does not affect ordinary holders: it corrupts only the pair, the single address whose balance sets the on-chain price. With the token reserve reduced to 1 wei, the pool grossly overvalues the token. The attacker then swaps against the mispriced pool and withdraws almost its entire paired reserve, for example the WBNB or stablecoin side. The loss thus falls on the liquidity providers, not on individual users. The entire sequence, from the triggering transfer to the final swap, executes atomically in one transaction.
 
Despite operating as independent projects, FFIST and Utopia share an identical implementation of the \texttt{\_airdrop} function. This indicates that one contract was forked from the other (or both from a common ancestor) without any security review of the airdrop logic. Had security auditors promptly utilized DeFiScreener to screen for analogous vulnerable functions immediately following the Utopia exploit, they could have leveraged a 7-hour time window to proactively intercept the subsequent attack on FFIST.

Due to space constraints, Case Study II (Serial Template Exploitation) is detailed in Appendix \ref{appendix:motivation}.

\subsection{Technical Insight}
Taken together, the two case studies demonstrate a consistent empirical pattern: a critical vulnerability introduced in a pioneer protocol is faithfully inherited by its derivatives, enabling attackers to re-exploit the same attack pattern against multiple targets, sometimes within the same day, as in the FFIST and Utopia incidents, and sometimes across a multi-month campaign, as in the reflection token series. Traditional reactive measures, such as post-mortem audits, often fail to prevent such exploits due to the sheer volume of forked protocols. 

This gap highlights the urgent need for a proactive pre-screening mechanism. By abstracting these historical exploits into structured attack cases, we can leverage historical case matching to identify identical or mutated vulnerability patterns in candidate contracts before deployment, or ensure that newly discovered attack patterns can be cross-referenced with deployed contracts, fundamentally shifting the defense paradigm from reactive patching to predictive prevention.

\begin{figure*}[!t]
  \includegraphics[width=7 in]{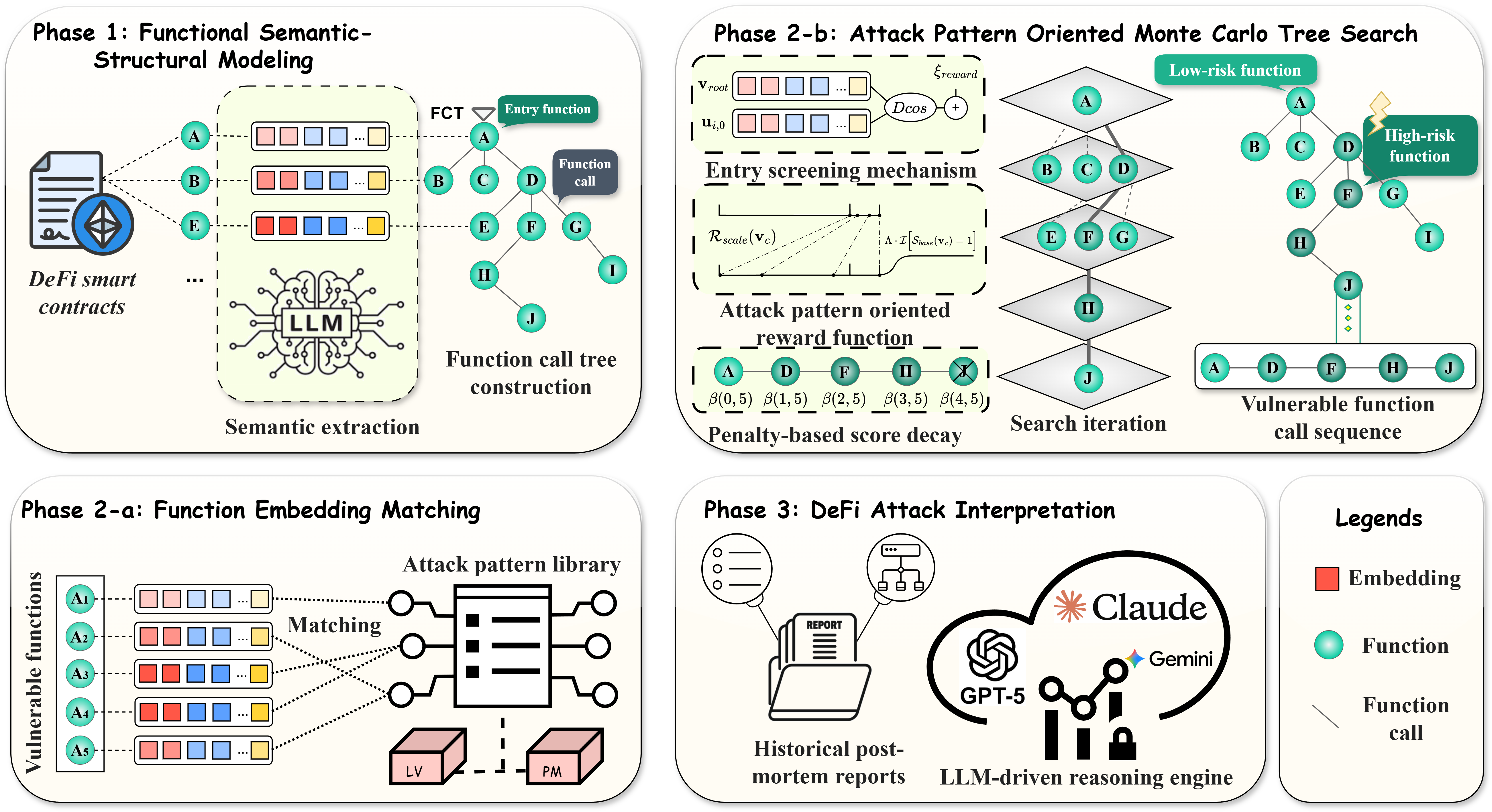}
  \caption{Overview of the DeFiScreener framework, a novel pre-screening approach that proactively identifies potentially vulnerable functions and call sequences at an early stage, enabling targeted and efficient downstream analysis.}
  \label{fig:DeFiScreener}
\end{figure*}

\section{DEFISCREENER}
In this section, we introduce DeFiScreener, which is depicted in Fig.\ref{fig:DeFiScreener}. DeFiScreener consists of three phases: functional semantic-structural modeling, vulnerable function and sequences screening and DeFi attack interpretation.

\subsection{Workflow of DeFiScreener}
\textbf{Method Overview.} We propose DeFiScreener to pre-screen vulnerable functions and its call sequences in DeFi smart contracts. At a high level, the three phases of DeFiScreener are: (1) In the \textit{functional semantic-structural modeling} phase, we leverage LLMs to extract the semantic features of each function within the contract, and construct function call trees based on the call dependencies derived from static code analysis.
(2) In the \textit{vulnerable functions and sequences screening} phase, we compute the similarity between the semantic embedding of each function and the established attack pattern embeddings to identify vulnerable functions. For the analysis of function call sequences, we introduce an entry screening mechanism designed to prioritize various functions as the initial node of MCTS. Furthermore, during the search process, our derived attack pattern oriented reward function, coupled with a penalty-based score decay algorithm, significantly enhances overall search precision.
(3) In the \textit{DeFi attack interpretation} phase, we provide explainable analyzes and preventative recommendations for the matched historical cases by utilizing LLMs, thereby facilitating the auditors' interpretation of the pre-screening results and their subsequent selection of dedicated vulnerability detectors.

In what follows, we elaborate on the three phases in detail.

\subsection{Functional Semantic-Structural Modeling}
In this phase, we leverage the LLM to extract the semantic embeddings of each function for matching and construct the \textsf{FCT} based on the function call dependencies derived from static code to serve as the input for the subsequent APO-MCTS.

\subsubsection{Semantic Extraction}
The representation of each function is central to pre-screening, because DeFi exploits arise from a function's financial and logical semantics rather than its surface syntax. Hand-crafted vulnerability features and expert rules bridge this gap poorly, because they demand per-attack-type engineering and generalize weakly to unseen patterns. Token, AST, or graph similarity, in turn, encodes form rather than intent and is easily broken by cosmetic edits. We therefore encode each function with a code-specialized LLM pretrained on large code corpora, capturing function-level intent. A single embedding space also covers diverse attack types without attack-specific rules or per-dataset recalibration, which underpins the plug-in Attack Pattern Library used in later phases.

Concretely, we instantiate the encoder with code-specialized LLMs and apply 4-bit NormalFloat (NF4) quantization together with double quantization~\cite{dettmers2023qlora}, reducing GPU memory overhead while preserving the fidelity of the semantic representation.

A function is mapped into a token sequence. To balance contextual coverage with computational tractability, $n$ is truncated to a predefined threshold. To mitigate numerical instability during inference, we adopt the BFloat16 format \cite{chowdhery2023palm} throughout the pipeline. 
Consequently, the LLM generates a sequence of context-aware hidden states, providing a fine-grained semantic representation of the function used for matching in the subsequent phases.

\subsubsection{Function Call Tree Construction}
DeFiScreener builds each \textsf{FCT} from the full static source of the target protocol, retrieved through the blockchain. When a retrieved contract is a proxy that delegates execution to a separate implementation, we collect that implementation contract rather than the near-empty proxy shell. A non-proxy contract is collected as is. All source files under the target address are then gathered into a single per-protocol codebase, from which the \textsf{FCT} is constructed. A call edge is added whenever the callee's implementation is present in this codebase, so intra-protocol calls and resolved proxy targets are both captured. A call whose target has no body in the codebase, such as an interface declaration or a cross-protocol call, terminates the corresponding branch. DeFiScreener is therefore intra-protocol in scope: it resolves calls within the collected protocol source, but does not follow external calls whose implementation lies outside it. This boundary is consistent with the interface-only misses reported in RQ1 and is discussed further in Section \ref{subsec:limitations}.

The topological structure of a \textsf{FCT} is governed by static call relationships, where nodes and edges encapsulate function embeddings and call behaviors, respectively. We model each function within a DeFi protocol as a discrete entry point (i.e., the root node) to generate a comprehensive set of \textsf{FCT}s, the cardinality of which equals the total number of functions. Within this framework, the parent-child hierarchy between nodes serves as a direct mapping of the underlying functional call logic.

We employ Depth-First Search (DFS) to generate call sequences. The rationale for selecting DFS lies in its backtracking capability; upon reaching a leaf node, the corresponding call path can be efficiently reconstructed through recursion or stack backtracking. Conversely, utilizing Breadth-First Search (BFS) for sequence generation poses significant overhead, as it fails to inherently preserve hierarchical ancestry. Specifically, unless parental pointers are explicitly maintained and each traversal step is logged during dequeuing, recovering the direct lineage of a leaf node remains computationally cumbersome.

Certain protocols contain cyclic function calls that form loops in the call graph, which could drive a naive traversal into infinite recursion~\cite{sun2024gptscan}. To guarantee termination, we track the nodes along the current path with an auxiliary vector and, upon revisiting a node, treat the recurrent node as a terminal leaf. This does not discard cyclic behavior, because a separate \textsf{FCT} is rooted at every function, so a cycle is unrolled across roots. The cycle $A\!\to\!B\!\to\!C\!\to\!D\!\to\!A$ yields the rooted paths $A$-$B$-$C$-$D$, $B$-$C$-$D$-$A$, $C$-$D$-$A$-$B$, and $D$-$A$-$B$-$C$, one per entry function. These are exactly the rotations an exploit could traverse depending on where it enters the cycle. The per-function \textsf{FCT} set therefore recovers the same reachable call sequences as a call graph, cutting only the redundant re-traversal that a cycle would otherwise repeat indefinitely. We thus adopt \textsf{FCT}s rather than a raw call graph, keeping the efficient path reconstruction of a tree without losing cyclic call behavior.

With the comprehensive set of \textsf{FCT}s constructed, we derive APO-MCTS to pinpoint extraction of vulnerable function call sequences within the vast call space.

\subsection{Vulnerable Functions and Sequences Screening}
In this phase, based on the function embeddings and \textsf{FCT}s modeled in Phase 1, we perform embedding matching on each function and apply APO-MCTS for the \textsf{FCT}s to capture vulnerable functions and sequences within the protocol.

\subsubsection{Function Embedding Matching}
Each function embedding is matched against those from disclosed attack incidents stored in the attack pattern library. 

\textbf{Attack Pattern Library (APL)} is the reference set against which each function embedding is matched, denoted $\mathcal{L}=\{P_1,\dots,P_N\}$. Each record $P_i$ is distilled from a single disclosed, on-chain-confirmed exploit, not from a whole contract or a set of arbitrary functions. It stores the semantic embeddings of the function or functions that carried the exploitable flaw, together with the ordered call sequence that the real attack transaction traversed to trigger it. In other words, the APL captures the semantic intent of the flawed function and the call structure that reaches and triggers that flaw in the \textsf{FCT}. It excludes the dynamic transaction context, which is runtime-dependent and lies outside our static scope. Matching a candidate against $P_i$ thus asks whether the contract contains the same flawed function and triggering call structure that an attacker has already weaponized. Accordingly, the stored record is the exploit path recovered from a real incident (semantics and structure of static code).

Given the APL, DeFiScreener matches the embedding of each candidate function against the stored patterns in $\mathcal{L}$. A function is flagged as vulnerable when its cosine distance to a stored malicious embedding falls below a predefined threshold, whose sensitivity we analyze in RQ4.




\subsubsection{Attack Pattern Oriented Monte Carlo Tree Search}
Up to now, a vast of function call sequences has been generated across multiple \textsf{FCT}s. The cardinality of the potential sequence space (calculated by the product of entry points and leaf nodes) can be prohibitively large, leading to a state-space explosion. Performing exhaustive deep matching on all candidate sequences would be computationally intractable. To mitigate this overhead, we propose APO-MCTS to identify the most vulnerable function call sequences.

During the implementation of APO-MCTS, we explore three pivotal optimization strategies to enhance the identification of vulnerable function call sequences: (1) \textbf{Entry Screening Mechanism.} Screening the most suspicious \textsf{FCT} to search. (2) \textbf{Attack Pattern Oriented Reward Function.} Enhancing the precision of exploring vulnerable child nodes. (3) \textbf{Penalty-based Score Decay Algorithm.} Mitigating the probability of executing duplicate branches.

\textbf{Entry screening mechanism.}
Prioritizing the \textsf{FCT} for exploration is critical, as the selection of the root node directly affects the exposure depth of vulnerable sequences. If the entry function already exhibits strong semantic relevance to known attack patterns, the target vulnerable sequence can often be recovered by exploring only the shallow regions of the \textsf{FCT}. Motivated by this observation, we design an entry screening mechanism that assigns a suspiciousness score, denoted as $\mathcal{E}(T)$, to each candidate \textsf{FCT} prior to the formal search. The $\mathcal{E}(T)$ comprises two components: a differential value derived from pattern matching against the attack pattern library, and a reward value attributed to the second-layer child nodes.

Formally, for a candidate \textsf{FCT} $T$ rooted at function $f_{root}$, we denote its semantic embedding as $\mathbf{v}_{root} = \Phi(f_{root})$, where $\Phi$ is the semantic embedding extraction function for function $f_{root}$. Let $\mathcal{L} = \{ P_1, P_2, \dots, P_N \}$ be the comprehensive APL, where $\mathbf{u}_{i,0} \in P_i$ serves as the entry point of that pattern. We first establish the differential value $\xi_{diff}$ by computing the minimum spatial discrepancy between the candidate entry and all observed attack entries using cosine distance $\mathcal{D}_{cos}$:

\begin{equation}
    \xi_{diff} = \alpha - \min_{P_i \in \mathcal{L}} \underbrace{\mathcal{D}_{cos}(\mathbf{v}_{root}, \mathbf{u}_{i,0})}_{\text{cosine distance}}, \mathcal{D}_{cos}:R^d \times R^d\to[0,2]
\end{equation}

where $\alpha$ is a predefined scaling constant bounding the maximum baseline suspiciousness. $\xi_{diff}$ effectively measures the semantic proximity of the root node to the closest known threat.

Furthermore, to enhance the screening sensitivity towards localized structural cues, we introduce a conditional reward strategy. When $f_{root}$ exactly matches the entry function of a specific historical attack pattern $P_k$ (i.e., $\mathcal{D}_{cos}(\mathbf{v}_{root}, \mathbf{u}_{k,0}) = 0$), we inspect the immediate second-layer descendants, denoted as $\mathcal{S}ec(f_{root})$. If any direct child node is semantically aligned with any functional component within the corresponding pattern $P_k$, an additional structural reward $\xi_{reward}$ is assigned. This logic is mathematically encapsulated by an indicator function $\mathcal{I}(\cdot)$:

\begin{equation}
    \begin{split}
        \xi_{reward} =& \theta \cdot \mathcal{I} \Big( \exists f_c \in \mathcal{S}ec(f_{root}), \exists \mathbf{u} \in P_k \\
        &\mid \mathcal{D}_{cos}(\Phi(f_c), \mathbf{u}) = 0 \Big)
    \end{split}
\end{equation}

where $\theta$ is the reward penalty coefficient. Consequently, the final suspiciousness score of an entry jointly reflects the root-level differential value and the second-layer reward $\mathcal{E}(T) = \xi_{diff} + \xi_{reward}$. We then rank all candidate \textsf{FCT}s in descending order according to this score and select the highest-ranked entry function as the root for MCTS. 

We restrict this reward to the entry and its direct children (the second layer) for two reasons. First, the entry score is only a lightweight pre-ranking that decides which \textsf{FCT} to search first. The complete vulnerable sequence, at any depth, is subsequently recovered by APO-MCTS. A shallow, local structural cue is therefore sufficient. Second, the call graph is stored as an adjacency map from each function to its direct children, so reading the second layer is a single lookup. Extending the reward to the third layer would require expanding every child as a new source and querying again, which multiplies the number of inspected nodes with depth. Because this score is computed for every candidate entry, that cost compounds across the whole protocol. The second layer thus supplies the minimal structural context needed to disambiguate promising entries at the lowest cost, while depth is delegated to the search.

\textbf{Attack Pattern Oriented Reward Function.}
In the exploration phase of APO-MCTS, designing an effective reward function is imperative to guide the search algorithm towards branches that exhibit a high probability of vulnerability. To this end, we propose an attack pattern oriented reward function that evaluates the semantic relevance of a candidate node against the established attack pattern library and applies a non-linear scaling mechanism to amplify highly suspicious traits.

For a given node with semantic embedding $\mathbf{v}_c$, we first compute its minimum cosine distance to all known malicious functional embeddings $\mathbf{u} \in \mathcal{L}$, mapping this distance to a normalized base similarity score $\mathcal{S}_{base} \in [0, 1]$:

\begin{equation}
    \mathcal{S}_{base}(\mathbf{v}_c) = 1 - \frac{1}{\alpha} \min_{\mathbf{u} \in \mathcal{L}} \mathcal{D}_{cos}(\mathbf{v}_c, \mathbf{u})
\end{equation}

To effectively differentiate highly similar functions from benign ones, we introduce a logarithmic barrier scaling mechanism. Linear similarity scores often fail to provide sufficient gradients when approaching the upper bound (i.e., highly suspicious nodes). By utilizing a logarithmic barrier, we heavily penalize deviations from perfect alignment while completely truncating low-relevance scores. Specifically, given a predefined truncation threshold $s_{min}$ and a marginal offset $\epsilon$ to prevent singularity and preserve precision at the extreme edge, the log-distance is defined as:

\begin{equation}
    d_{log}(x) = \log_{10}(1 + \epsilon - x)
\end{equation}
The scaled reward $\mathcal{R}_{scale}$ is subsequently computed via linear interpolation of the log-distances:

\begin{equation}
    \mathcal{R}_{scale}(\mathbf{v}_c) = \begin{cases} 0 & \text{if } \mathcal{S}_{base}(\mathbf{v}_c) \le s_{min}  \\ \frac{d_{log}(\mathcal{S}_{base}) - d_{log}(s_{min})}{d_{log}(1.0) - d_{log}(s_{min})} & s_{min} < \mathcal{S}_{base}(\mathbf{v}_c) < 1.0 \\ 1 & \text{if } \mathcal{S}_{base}(\mathbf{v}_c) \ge 1.0 \end{cases} 
\end{equation}

This continuous mapping ensures that as the similarity $\mathcal{S}_{base}$ marginally increases toward $1.0$, the resulting reward grows exponentially, driving the APO-MCTS to greedily exploit these promising paths. Furthermore, to maximally incentivize exact semantic matches, we apply a substantial reward multiplier $\Lambda$ when the minimum cosine distance is strictly zero (i.e., $\mathcal{D}_{cos} = 0$). The final evaluation reward $\mathcal{R}(\mathbf{v}_c)$ is formalized as:

\begin{equation}
    \mathcal{R}(\mathbf{v}_c) = \mathcal{R}_{scale}(\mathbf{v}_c) \cdot \Big( 1 + \underbrace{\Lambda \cdot \mathcal{I} \big[ \mathcal{S}_{base}(\mathbf{v}_c) = 1 \big]}_{\text{additional reward score}} \Big)
\end{equation}

Ultimately, the APO-MCTS algorithm aggregates the features of the current node and its potential valid actions, calculating the respective rewards. The algorithm selects the maximum reward value among these features to dynamically update the tree policy, thereby ensuring that the search trajectory is aggressively directed toward the most semantically vulnerable execution sequences.

\textbf{Penalty-based Score Decay Algorithm.}
Standard MCTS balances exploration and exploitation through the upper confidence bound applied to trees. However, within the context of vulnerability screening, vulnerable function call sequences often yield exceptionally high rewards. This can induce a pathological exploitation bias, causing the MCTS to repeatedly converge on identical sequence branches across multiple execution rounds, thereby limiting the discovery of diverse vulnerable execution paths.

To systematically mitigate branch duplication and encourage lateral exploration, we introduce a Penalty-based Score Decay Algorithm. This mechanism retroactively depreciates the accumulated rewards of previously extracted function call sequences, forcing the search tree to diversify its trajectory.

Let $\mathcal{H}$ denote the set of historical best sequences extracted in prior iterations. Suppose a historical sequence $Seq \in \mathcal{H}$ of length $L$ is represented as a sequence of nodes $Seq = (n_0, n_1, \dots, n_{L-1})$, where $n_d$ denotes the node at topological depth $d$. Upon the completion of a search iteration, we traverse the \textsf{FCT} along the path of $Seq$ and apply a depth-aware multiplicative decay factor $\beta(d, L)$ to the total accumulated reward $\mathcal{R}_{total}(n_d)$ of each matched node. The node's accumulated reward is subsequently updated via:
\begin{equation}
    \mathcal{R}'_{total}(n_d) = \underbrace{\mathcal{R}_{total}(n_d)}_{\text{total rewards}} \cdot \underbrace{\beta(d, L)}_{\text{decay factor}}, \beta(d, L) = 1 - \frac{1}{2^{L - 1 - d}}
\end{equation}
The decay factor $\beta(d, L)$ is formulated as a function of the node's distance to the terminal leaf of the sequence.

This hierarchical penalization strategy exhibits two critical properties:
\begin{itemize}
\item Terminal annihilation: At the leaf node of the historical sequence ($d = L - 1$), the distance is zero, resulting in $\beta = 0$. The reward is completely neutralized ($\mathcal{R}'_{total} = 0$), effectively truncating this exact path from future final selections.
\item Prefix preservation: For nodes situated closer to the root ($d \to 0$), the penalty undergoes exponential decay (i.e., $\beta \to 1$).
\end{itemize}

By preserving the majority of the reward for highly suspicious prefix paths while severely penalizing the exact leaf formations, this algorithm ensures that the APO-MCTS retains the intelligence of vulnerable entry points but is strictly coerced into exploring unvisited sibling branches in subsequent function calls.

The APO-MCTS iterations are detailed in Appendix \ref{appendix:APO-MCTS}.

Upon the conclusion of each search, we extract the sequence with the highest suspiciousness score and apply a discount factor $\gamma$ to the maximum score $\xi_{Max}$ of the selected entrance. This penalty mechanism effectively diminishes the selection probability of the corresponding entry function in subsequent iterations, thereby preventing the search process from converging prematurely on localized high-scoring paths and ensuring a broader coverage of the potential attack space.

\subsection{DeFi Attack Interpretation}
The interpretability of DeFi attack detection has long been marginalized by existing automated tools, which frequently yield binary detection results devoid of actionable context. To bridge this semantic gap, our framework systematically correlates matched attack events with their underlying root causes, which are distilled from an extensive corpus of historical post-mortem reports and expert security audits.

To synthesize this structured mapping into human-readable threat intelligence, DeFiScreener integrates a LLM-driven reasoning engine. Operating under strict, blockchain-security-specific system instructions, the reasoning engine processes the mapped sequences and historical data to generate a multi-dimensional security incident report. This automated report provides security analysts with a concise attack conclusion, a chronologically reconstructed attack path, and a rigorous type-to-cause risk mapping substantiated by sequence-level evidence. Furthermore, the engine delivers targeted, high-value mitigation strategies while explicitly isolating analytical uncertainties, ensuring that the generated insights remain strictly bounded by empirical evidence rather than unverified heuristic assumptions.
The prompt template and corresponding sample case for the interpretations are provided in our GitHub repository.

\section{EVALUATION AND ANALYSIS}
In this section, we aim to answer the following research questions (RQs):

\textbf{RQ1:} How effectively does DeFiScreener screen for DeFi attacks?

\textbf{RQ2:} How well does DeFiScreener generalize to previously unseen (zero-day) attacks?

\textbf{RQ3:} To what extent does DeFiScreener identify vulnerable functions or call sequences overlooked by SOTA tools?

\textbf{RQ4:} How do the different phases contribute to the overall effectiveness of DeFiScreener?

\textbf{Dataset.} Our experiments are conducted on three public datasets. As shown below, Dataset \uppercase\expandafter{\romannumeral 1} is constructed in this work from publicly disclosed incidents and supports RQ1, RQ2, and RQ4. Dataset \uppercase\expandafter{\romannumeral 2} and Dataset \uppercase\expandafter{\romannumeral 3} are public benchmarks from \cite{zhong2025detecting} and \cite{wen2024foray}, respectively, and their intersection with Dataset \uppercase\expandafter{\romannumeral 1} supports the downstream evaluation in RQ3.

\begin{itemize}
    \item Dataset \uppercase\expandafter{\romannumeral 1}. We construct Dataset \uppercase\expandafter{\romannumeral 1} from the disclosed DeFi attack incidents catalogued on SlowMist~\cite{slowmist} and DeFiHackLabs~\cite{DeFiHackLabs}, publicly disclosed between 2017 and 2025. This pool overlaps substantially with the 573 incidents, but applies the three stricter, reproducibility-oriented inclusion criteria below: (i) it corresponds to a real, on-chain-confirmed attack whose root cause is documented in a public post-mortem or audit report; (ii) the victim contract's Solidity source code is verified and retrievable through a block explorer; and (iii) the vulnerable function and its call sequence can be located in that source and labeled from the report. Applying these criteria yields 224 candidate incidents that span the two broad attack categories introduced in Section~\ref{sec:defi-attacks}: business logic vulnerabilities, and price and oracle manipulations. These two categories subsume the 26 specific attack types listed in Table~\ref{tab:RQ1-1}. We then remove 17 of the 224 candidates for two reasons. First, in four incidents~\cite{ODOS, YIEDL, SwarmMarkets, XSURGE}, the vulnerable function is only an interface declaration, and its implementation lies outside the project's local codebase. Second, in thirteen incidents~\cite{EXcommunity, Miner, FloorProtocol, GoodDollar, APEDAO, BCT, BelugaDex, Audius, WiseLending, Platypus, Palmswap, LFI-Token, NUM}, the vulnerable function reported in prior case studies is absent from the retrievable code, typically because an upgradable proxy had changed the deployed logic. The remaining 207 incidents comprise 16,775 functions across 26 attack types, with an aggregate loss of about \$2.97 billion, and form the evaluation set summarized in Table~\ref{tab:RQ1-1}. For each retained incident, we independently labeled its vulnerable functions or the corresponding call sequences, using the public post-mortem and audit reports as the reference standard.

    \item Dataset \uppercase\expandafter{\romannumeral 2}. To evaluate the effectiveness of the DeFiScreener pre-screening mechanism, we employed the benchmark dataset constructed by \cite{zhong2025detecting}. This dataset encompasses a comprehensive collection of real-world price manipulation attacks occurring within the Ethereum and Binance Smart Chain (BSC) ecosystems. The data collection methodology cross-validates and integrates records from three independent sources: 54 attack incidents identified by DeFort \cite{xie2024defort}, 55 security events documented in the DeFiHackLabs \cite{DeFiHackLabs} open-source vulnerability repository, and 31 attacks provided and verified by industry partners.
    \item Dataset \uppercase\expandafter{\romannumeral 3}. This dataset comprises 34 benchmark cases of logical vulnerabilities in DeFi smart contracts, curated from the DeFiHackLabs \cite{DeFiHackLabs} repository and spanning real-world exploit incidents recorded between January 2022 and July 2023. Each benchmark corresponds to a verified on-chain attack event and is classified into one of four logical flaw categories: Token Burn, Pump \& Dump, Price Discrepancy, and Swap Rate Manipulation. The exploits represented in this dataset resulted in documented financial losses exceeding USD 21 million, providing a rigorous and representative benchmark suite for the evaluation of DeFi-oriented logical vulnerability detection systems.
\end{itemize}

\textbf{Implementation.} DeFiScreener accepts smart contract source code, retrieved via the Web3 API, as its primary input. DeFiScreener outputs the identified vulnerable functions or sequences, accompanied by comprehensive explanatory analyses. For the function embedding extraction in phase 1, we employ Codegemma\cite{Codegemma}, Codellama\cite{Codellama}, Deepseek-coder\cite{Deepseek-coder}, Qwen2.5-coder\cite{Qwen2.5-coder}, and Starcoder2\cite{Starcoder2}. We set the reward penalty coefficient $\theta$ to 5, the scaling constant $\alpha$ to 2, and the truncation threshold $s_{min}$ to 0.8.

Currently, DeFiScreener supports the analysis of smart contracts deployed across multiple blockchains, including Arbitrum \cite{arbiscan}, Base \cite{basescan}, BNB Chain \cite{bscscan}, Ethereum \cite{etherscan}, and Polygon \cite{polygonscan}. All experimental evaluations are conducted on a workstation equipped with an AMD Ryzen 9 8945HX with Radeon Graphics (2.50 GHz), an NVIDIA GeForce RTX 5060 Ti GPU (8GB VRAM), and 16GB of RAM.

\subsection{RQ1: Pre-screening Effectiveness}
In this RQ, we evaluate the pre-screening capability of DeFiScreener. Specifically, this evaluation comprises three primary aspects: the overall screening effectiveness, the effectiveness of vulnerable function matching, and the effectiveness of screening vulnerable function call sequences.

\begin{table}[htbp]
    \centering
    \caption{The overall screening capability of DeFiScreener. Here, types are grouped under the two broad categories (logic vulnerabilities above, price manipulations below), \#TP denotes the number of successfully screened protocols, \#FP represents the number of functions triggering false positives, and \#ALL indicates the total number of evaluated protocols.}
    
    
    \begin{tabular}{l | c | c c | c}
        \hline
        \textbf{Attack Type} & \textbf{Loss(\$)} & \#TP & \#FP & \#ALL  \\
        \hline
        
        Access Control & 1085.33M & 25 & 5 & 25 \\
        Misconfiguration & 92K & 2 & 0 & 2 \\
        Incorrect logic & 14.54M & 11 & 5 & 11 \\
        Reentrancy & 37.54M & 16 & 4 & 17 \\  
        Insufficient validation & 287.77M & 26  & 7 & 27 \\ 
        Unchecked User Input & 7.1M & 2 & 0 & 2 \\
        Business Logic Flaw & 50.13M & 16 & 5 & 16 \\ 
        Arbitrary call & 2.6M & 8 & 2 & 8 \\
        Math  & 7.01M & 2 & 1 & 2 \\
        Precision  & 77.81M & 7 & 0 & 7 \\
        Bad randomness & 236.93K & 4 & 0 & 4 \\
        Miscalculation & 7.78M & 7 & 2 & 8 \\
        Deflationary token & 3.78M & 9 & 1 & 9 \\
        skim & 46.07K & 1 & 0 & 1 \\
        Signature & 3.02M & 2 & 0 & 2  \\
        DoS & 34M & 1 & 1 & 1 \\
        CrossChain & 81.97M & 3 & 0 & 3 \\
        Lack of permission control & 442K & 1 & 0 & 1 \\
        \hline
        
        Price Manipulation & 49.38M & 18 & 0 & 18 \\
        Flashloans & 171.29M & 20 & 4 & 20 \\ 
        ERC20 & 954.99M & 15 & 1 & 15 \\ 
        Arbitrage & 21.88K & 2 & 0 & 2 \\
        Slippage & 135K & 1 & 0 & 1 \\
        Oracle & 88.08M & 3 & 0 & 3 \\
        Rug pull  & 218.13K & 1 & 0 & 1 \\
        MaliciousProposal & 3M & 1 & 0 & 1 \\
        \hline
        \textbf{Total} & \textbf{2968.31M} & \textbf{204} & \textbf{38} & \textbf{207} \\
        \hline
    \end{tabular}
    \label{tab:RQ1-1}
\end{table}

\begin{table*}[t]
    \centering
    \caption{Screening results of vulnerable sequences for DeFiScreener. Here, $C_N$, $F_N$, and $S_N$ denote the total count of contracts, functions, and sequences, respectively. GT denotes the number of ground-truth vulnerable sequences for that protocol. TP indicates the complete capture of vulnerable sequences, MP represents the partial extraction of vulnerable fragments, and FN signifies the complete failure to identify any vulnerable sequences.}
    \begin{tabularx}{\textwidth}{c l c c c c c c c | c l c c c c c c c }
        \toprule 
        \textbf{\#} & \textbf{Protocol} & \textbf{$C_N$} & \textbf{$F_N$} & \textbf{$S_N$} & GT & TP & MP & FN & \textbf{\#} & \textbf{Protocol} & \textbf{$C_N$} & \textbf{$F_N$} & \textbf{$S_N$} & GT & TP & MP & FN \\
        \midrule 
        1 & AES & 1 & 41 & 59 & 1 & 1 & 0 & 0 & 25 & Mosca & 1 & 34 & 19 & 1 & 1 & 0 & 0 \\
        2 & ApeCoin & 16 & 61 & 65 & 1 & 1 & 0 & 0 & 26 & Omni NFT & 56 & 240 & 1532 & 1 & 1 & 0 & 0 \\
        3 & AstridProtocol & 32 & 130 & 92 & 1 & 1 & 0 & 0 & 27 & OrbitChain & 1 & 40 & 36 & 1 & 1 & 0 & 0 \\
        4 & Auctus & 4 & 7 & 7 & 1 & 1 & 0 & 0 & 28 & PancakeBunny & 37 & 125 & 545 & 1 & 1 & 0 & 0 \\
        5 & BabySwap & 22 & 109 & 494 & 1 & 1 & 0 & 0 & 29 & PancakeHunny & 1 & 176 & 821 & 1 & 0 & 1 & 0 \\
        6 & BedRock Defi & 22 & 93 & 58 & 1 & 1 & 0 & 0 & 30 & Paraluni & 1 & 138 & 2741 & 1 & 1 & 0 & 0 \\
        7 & Burner & 1 & 49 & 46 & 2 & 2 & 0 & 0 & 31 & PredyFinance & 60 & 283 & 181 & 1 & 1 & 0 & 0 \\
        8 & BXH & 1 & 95 & 593 & 1 & 1 & 0 & 0 & 32 & QTN & 1 & 30 & 32 & 2 & 2 & 0 & 0 \\
        9 & CloberDEX & 29 & 151 & 65 & 1 & 0 & 1 & 0 & 33 & Qubit Finance & 16 & 101 & 57 & 1 & 1 & 0 & 0 \\
        10 & Cover Protocol & 15 & 67 & 151 & 3 & 2 & 1 & 0 & 34 & Rari Capital & 34 & 212 & 979 & 1 & 1 & 0 & 0 \\
        11 & CowSwap & 16 & 46 & 42 & 1 & 1 & 0 & 0 & 35 & RES token & 1 & 53 & 58 & 1 & 1 & 0 & 0 \\
        12 & CS Token & 1 & 85 & 65 & 1 & 1 & 0 & 0 & 36 & RL Token & 12 & 54 & 27 & 1 & 1 & 0 & 0 \\
        13 & DPC & 1 & 80 & 299 & 1 & 1 & 0 & 0 & 37 & RoeFinance & 98 & 313 & 2123 & 1 & 1 & 0 & 0 \\
        14 & EFLeverVault & 1 & 35 & 135 & 1 & 1 & 0 & 0 & 38 & RuggedArt & 21 & 76 & 41 & 1 & 1 & 0 & 0 \\
        15 & EGD Finance & 7 & 42 & 19 & 1 & 1 & 0 & 0 & 39 & SafeDollar & 2 & 96 & 435 & 1 & 1 & 0 & 0 \\
        16 & H2O & 1 & 23 & 18 & 1 & 1 & 0 & 0 & 40 & ShibaToken & 21 & 90 & 66 & 2 & 2 & 0 & 0 \\
        17 & Hegic Options & 47 & 130 & 169 & 1 & 1 & 0 & 0 & 41 & SorStaking & 11 & 43 & 27 & 2 & 2 & 0 & 0 \\
        18 & JAY & 1 & 54 & 65 & 1 & 1 & 0 & 0 & 42 & TCH & 1 & 59 & 32 & 1 & 1 & 0 & 0 \\
        19 & Juice & 7 & 29 & 29 & 1 & 1 & 0 & 0 & 43 & Thunder Brawl & 2 & 73 & 95 & 1 & 1 & 0 & 0 \\
        20 & Lodestar & 4 & 7 & 2 & 1 & 1 & 0 & 0 & 44 & TreasureDAO & 18 & 39 & 31 & 1 & 0 & 1 & 0 \\
        21 & MBC & 1 & 50 & 50 & 1 & 1 & 0 & 0 & 45 & WECO & 1 & 26 & 26 & 1 & 1 & 0 & 0 \\
        22 & MIMSpell$_2$ & 14 & 57 & 68 & 2 & 2 & 0 & 0 & 46 & wsm & 1 & 55 & 25 & 1 & 1 & 0 & 0 \\
        23 & MixedSwapRouter & 29 & 110 & 158 & 1 & 1 & 0 & 0 & 47 & ZABU Finance & 6 & 118 & 691 & 2 & 2 & 0 & 0 \\
        24 & MonoX Finance & 20 & 96 & 361 & 1 & 1 & 0 & 0 &  & \textbf{Total} & \textbf{696} & \textbf{4121} & \textbf{13730} & \textbf{55} & \textbf{51} & \textbf{4} & \textbf{0} \\
        \bottomrule
    \end{tabularx}
    \label{tab:RQ1-2}
\end{table*}

\subsubsection{The Overall Screening Effectiveness}
Table \ref{tab:RQ1-1} provides a detailed breakdown of the screening results generated by DeFiScreener across various attack types. We categorize the 207 analyzed security incidents into logic vulnerability attacks (e.g., \textit{Access Control}, \textit{Reentrancy}) and price manipulation attacks (e.g., \textit{Flashloans}, \textit{Rug pull}), and report their corresponding financial impact.

The analysis of the empirical data reveals the profound financial implications of the evaluated attack incidents, culminating in an aggregate loss of approximately \$2.97 billion (2,968.31M). Among the categorized attack types, vulnerabilities rooted in \textit{Access Control} and \textit{ERC20} standards inflicted the most substantial economic damages, amounting to \$1,085.33M and \$954.99M, respectively. In terms of occurrence frequency, \textit{Insufficient validation} (ALL=27), \textit{Access Control} (ALL=25), and \textit{Flashloans} (ALL=20) emerged as the most prevalent attack types.

Among a total of 207 real-world attack incidents, DeFiScreener effectively identified 204 incidents, achieving 98.55\% recall. The CloberDEX \cite{CloberDEX}, PancakeHunny \cite{PancakeHunny}, and TreasureDAO \cite{TreasureDAO} incidents are not counted as identified, because every one of their ground-truth sequences was only partially captured (MP in Table~\ref{tab:RQ1-2}), never fully. A detailed root-cause analysis of these three cases will be elaborated upon in the section of the effectiveness of screening vulnerable function call sequences.

During the evaluation, DeFiScreener generated 38 false positives, yielding 84.30\% precision. These FPs are predominantly concentrated in attack types such as \textit{Insufficient validation} and \textit{Access Control}. This phenomenon is primarily attributed to the presence of benign functions within the protocol source code that share similar semantic profiles with the targeted vulnerable functions, inadvertently leading to redundant matches during the vulnerable function matching. Nevertheless, such occurrences are statistically infrequent (accounting for only 38 out of 16,775 functions) and can be effectively mitigated by tightening the matching thresholds. Notably, across the 26 evaluated attack types, 15 types (e.g., Precision, CrossChain, and Oracle) achieved a zero false-positive rate.

\begin{table*}[t]
  \centering
  \small
  \setlength{\tabcolsep}{3.3pt}
  
  \caption{Evaluation of DeFiScreener under Simulated Zero-Day Attack Conditions. The protocols on the left denote existing data within the attack pattern library, whereas those on the right represent the target protocols under evaluation. For a detailed description of attack types, please refer to our GitHub repository.}

  \begin{tabularx}{\textwidth}{c | l l c c | l l c c | c c c}
  \toprule 
  \textbf{\#} & \textbf{Protocol} & \textbf{Date} & \textbf{Loss} & \textbf{Chain} & \textbf{Protocol} & \textbf{Date} & \textbf{Loss} & \textbf{Chain} & \textbf{Type} & \textbf{Function} & \textbf{Result} \\
  \midrule 
  
  1 & PancakeBunny & 2021/05/19 & 45M & BSC & PancakeHunny & 2024/06/03 & 1.93M & BSC & PM-1 & getReward & $\bullet$ \\

  2 & Rari Capital & 2022/04/30 & 80M & ETH & HundredFinance & 2023/04/15 & 7M & ETH & RE & borrowFresh & \ding{51} \\
  
  3 & NOVO Protocol & 2022/05/29 & 153K & BSC & ROI & 2022/09/28 & 44K & BSC & PM-2 & balanceOf & \ding{51} \\
  
  4 & NOVO Protocol & 2022/05/29 & 153K & BSC & BEVO & 2023/01/30 & 92K & BSC & PM-2 & balanceOf & \ding{51} \\
  
  5 & NOVO Protocol & 2022/05/29 & 153K & BSC & FDP Token & 2023/02/07 & 10K & BSC & PM-2 & balanceOf & \ding{51} \\
  
  6 & NOVO Protocol & 2022/05/29 & 153K & BSC & Sheep Token & 2023/02/10 & 3K & BSC & PM-2 & balanceOf & \ding{51} \\

  7 & Omni NFT & 2022/07/10 & 1.4M & ETH & CloberDEX & 2024/12/10 & 501K & BASE & RE & burn & \ding{55} \\
  
  8 & SpaceGodzilla & 2022/07/13 & 26K & BSC & MBC & 2022/11/29 & 5.6K & BSC & AC & swapAndLiquifyStepv1 & \ding{51} \\

  9 & OlympusDAO & 2022/10/21 & 292K & ETH & YIEDL & 2024/04/24 & 150K & ARB & LF & redeem & \ding{55} \\
  
  10 & SheepFarm & 2022/11/16 & 0.6K & BSC & FAPEN & 2023/05/29 & 0.6K & BSC & LF-2 & register & \ding{51} \\
  
  11 & TomInu Token & 2023/01/26 & 51K & ETH & BEVO & 2023/01/30 & 92K & BSC & PM-2 & deliver & \ding{51} \\
  
  12 & TomInu Token & 2023/01/26 & 51K & ETH & FDP Token & 2023/02/07 & 10K & BSC & PM-2 & deliver & \ding{51} \\

  13 & TomInu Token & 2023/01/26 & 51K & ETH & OLIFE & 2023/04/19 & 20K & BSC & PM-2 & deliver & \ding{51} \\
  
  14 & Utopia & 2023/07/20 & 119K & BSC & FFIST & 2023/07/20 & 110K & BSC & RD & \_airdrop & \ding{51} \\

  15 & Uwerx & 2023/08/02 & 300K & ETH & MetaDragon & 2024/05/29 & 180K & BSC & PM-3 & transfer & \ding{51} \\
  
  16 & Hopelend & 2023/10/18 & 110K & ETH & MahaLend & 2023/11/11 & 119K & ETH & MATH & rayDiv & \ding{51} \\
  
  17 & TIME & 2023/12/06 & 144K & ETH & HNet & 2023/12/07 & 1.3K & BSC & PM-3 & multicall & \ding{51} \\

  \bottomrule
  \end{tabularx}
  \label{tab:RQ3-1}
\end{table*}

\begin{table*}[t]
  \centering
  
  \caption{Comparison of results between running downstream tools independently and incorporating pre-screening, in price manipulation attack. The date indicates the attack date measured in UTC. The loss amount is in USD. DR refers to DeFiRanger \cite{wu2023defiranger},DT refers to DeFiTainter \cite{kong2023defitainter}, DF refers to DeFort \cite{xie2024defort}, DS refers to DeFiScope \cite{zhong2025detecting}, DSc refers to DeFiScreener. \ding{51} marks an incident it identifies correctly, \ding{55} marks a miss, and $\uparrow$ marks an incident it misses alone but discovers once DeFiScreener pre-screens ahead of it. UP reports, per incident, how many detectors are uplifted from a miss to a hit in this way.}

  \begin{tabularx}{\textwidth}{c l l c c c c c c c c c c c}
  \toprule 
  \textbf{\#} & \textbf{Protocol} & \textbf{Date} & \textbf{Loss} & \textbf{Chain} & \textbf{DR} & \textbf{DSc+DR} & \textbf{DT} & \textbf{DSc+DT} & \textbf{DF} & \textbf{DSc+DF} & \textbf{DS} & \textbf{DSc+DS} & \textbf{UP($\uparrow$)} \\
  \midrule

  1 & Harvest Finance & 2020/10/26 & 21.5M & ETH & \ding{51} & \ding{51} & \ding{51} & \ding{51} & \ding{51} & \ding{51} & \ding{55} & $\uparrow$ & \textbf{1} \\

2 & PancakeBunny & 2021/05/19 & 45M & BSC & \ding{55} & $\uparrow$ & \ding{51} & \ding{51} & \ding{51} & \ding{51} & \ding{51} & \ding{51} & \textbf{1} \\

3 & PancakeHunny & 2021/10/20 & 1.93M & BSC & \ding{55} & \ding{55} & \ding{55} & \ding{55} & \ding{55} & \ding{55} & \ding{55} & \ding{55} & \textbf{0} \\

4 & Cream Finance & 2021/10/27 & 130M & ETH & \ding{55} & $\uparrow$ & \ding{51} & \ding{51} & \ding{51} & \ding{51} & \ding{51} & \ding{51} & \textbf{1} \\

5 & MonoX Finance & 2021/11/30 & 31M & ETH & \ding{51} & \ding{51} & \ding{55} & $\uparrow$ & \ding{55} & $\uparrow$ & \ding{51} & \ding{51} & \textbf{2} \\

6 & InverseFinance & 2022/04/02 & 15.6M & ETH & \ding{55} & $\uparrow$ & \ding{55} & $\uparrow$ & \ding{55} & $\uparrow$ & \ding{55} & $\uparrow$ & \textbf{4} \\

7 & Wiener DOGE & 2022/04/25 & 30K & BSC & \ding{51} & \ding{51} & \ding{51} & \ding{51} & \ding{51} & \ding{51} & \ding{51} & \ding{51} & \textbf{0} \\

8 & HackerDao & 2022/05/24 & 65K & BSC & \ding{51} & \ding{51} & \ding{55} & $\uparrow$ & \ding{51} & \ding{51} & \ding{51} & \ding{51} & \textbf{1} \\

9 & BabyDoge & 2023/05/28 & 137K & BSC & \ding{55} & $\uparrow$ & \ding{55} & $\uparrow$ & \ding{55} & $\uparrow$ & \ding{51} & \ding{51} & \textbf{4} \\

10 & NOVO Protocol & 2022/05/29 & 76K & BSC & \ding{51} & \ding{51} & \ding{55} & $\uparrow$ & \ding{55} & $\uparrow$ & \ding{51} & \ding{51} & \textbf{2} \\

11 & Discover & 2022/06/06 & 11K & BSC & \ding{55} & $\uparrow$ & \ding{55} & $\uparrow$ & \ding{55} & $\uparrow$ & \ding{51} & \ding{51} & \textbf{4} \\

12 & SpaceGodzilla & 2022/07/13 & 25K & BSC & \ding{51} & \ding{51} & \ding{55} & $\uparrow$ & \ding{51} & \ding{51} & \ding{51} & \ding{51} & \textbf{1} \\

13 & EGD Finance & 2022/08/07 & 36K & BSC & \ding{55} & $\uparrow$ & \ding{51} & \ding{51} & \ding{51} & \ding{51} & \ding{51} & \ding{51} & \textbf{1} \\

14 & XSTABLE Protocol & 2022/08/09 & 56k & ETH & \ding{55} & $\uparrow$ & \ding{55} & $\uparrow$ & \ding{51} & \ding{51} & \ding{51} & \ding{51} & \textbf{2} \\

15 & BXH & 2022/09/28 & 40K & BSC & \ding{55} & $\uparrow$ & \ding{51} & \ding{51} & \ding{51} & \ding{51} & \ding{51} & \ding{51} & \textbf{1} \\

16 & ATK & 2022/10/12 & 61K & BSC & \ding{55} & $\uparrow$ & \ding{55} & $\uparrow$ & \ding{51} & \ding{51} & \ding{55} & $\uparrow$ & \textbf{3} \\

17 & BDEX & 2022/10/30 & 3K & BSC & \ding{55} & $\uparrow$ & \ding{51} & \ding{51} & \ding{51} & \ding{51} & \ding{55} & $\uparrow$ & \textbf{2} \\

18 & BBOX & 2022/11/16 & 12K & BSC & \ding{55} & $\uparrow$ & \ding{55} & $\uparrow$ & \ding{51} & \ding{51} & \ding{51} & \ding{51} & \textbf{2} \\

19 & MBC & 2022/11/29 & 5.9K & BSC & \ding{55} & $\uparrow$ & \ding{51} & \ding{51} & \ding{51} & \ding{51} & \ding{51} & \ding{51} & \textbf{1} \\

20 & AES & 2022/12/07 & 60K & BSC & \ding{55} & $\uparrow$ & \ding{55} & $\uparrow$ & \ding{51} & \ding{51} & \ding{51} & \ding{51} & \textbf{2} \\

21 & BGLD & 2022/12/12 & 18K & BSC & \ding{55} & $\uparrow$ & \ding{51} & \ding{51} & \ding{51} & \ding{51} & \ding{51} & \ding{51} & \textbf{1} \\

22 & RoeFinance & 2023/01/11 & 80K & ETH & \ding{55} & $\uparrow$ & \ding{55} & $\uparrow$ & \ding{55} & $\uparrow$ & \ding{51} & \ding{51} & \textbf{4} \\

23 & UPS & 2023/01/18 & 45K & ETH & \ding{51} & \ding{51} & \ding{51} & \ding{51} & \ding{51} & \ding{51} & \ding{51} & \ding{51} & \textbf{0} \\

24 & Starlink & 2023/02/16 & 12K & BSC & \ding{51} & \ding{51} & \ding{55} & $\uparrow$ & \ding{51} & \ding{51} & \ding{51} & \ding{51} & \textbf{1} \\

25 & swapX & 2023/02/27 & 1M & BSC & \ding{55} & $\uparrow$ & \ding{55} & $\uparrow$ & \ding{51} & \ding{51} & \ding{55} & $\uparrow$ & \textbf{4} \\

26 & SellToken01 & 2023/05/13 & 197k & BSC & \ding{51} & \ding{51} & \ding{55} & $\uparrow$ & \ding{55} & $\uparrow$ & \ding{51} & \ding{51} & \textbf{2} \\

27 & CS Token & 2023/05/23 & 714K & BSC & \ding{51} & \ding{51} & \ding{51} & \ding{51} & \ding{55} & $\uparrow$ & \ding{51} & \ding{51} & \textbf{1} \\

28 & Sturdy Finance & 2023/06/12 & 800K & ETH & \ding{55} & $\uparrow$ & \ding{55} & $\uparrow$ & \ding{55} & $\uparrow$ & \ding{51} & \ding{51} & \textbf{4} \\

29 & APEDAO & 2023/07/18 & 7K & BSC & \ding{51} & \textbf{-} & \ding{51} & \textbf{-} & \ding{55} & \ding{55} & \ding{51} & \textbf{-} & \textbf{0} \\

30 & FFIST & 2023/07/19 & 91K & BSC & \ding{55} & $\uparrow$ & \ding{55} & $\uparrow$ & \ding{51} & \ding{51} & \ding{51} & \ding{51} & \textbf{2} \\

31 & Uwerx & 2023/08/02 & 324K & ETH & \ding{51} & \ding{51} & \ding{55} & $\uparrow$ & \ding{55} & $\uparrow$ & \ding{51} & \ding{51} & \textbf{2} \\

32 & ZunamiProtocol & 2023/08/13 & 2M & ETH & \ding{55} & $\uparrow$ & \ding{55} & $\uparrow$ & \ding{55} & $\uparrow$ & \ding{51} & \ding{51} & \textbf{4} \\

33 & BFCToken & 2023/09/09 & 38K & BSC & \ding{51} & \ding{51} & \ding{55} & $\uparrow$ & \ding{55} & $\uparrow$ & \ding{51} & \ding{51} & \textbf{2} \\

34 & SATX & 2024/04/16 & 29K & BSC & \ding{51} & \ding{51} & \ding{55} & $\uparrow$ & \ding{55} & $\uparrow$ & \ding{51} & \ding{51} & \textbf{2} \\
  
  \midrule
  
  \# & Total (\ding{51} + $\uparrow$) & \textbf{-} & \textbf{-} & \textbf{-} & \textbf{14} & \textbf{(14 + 19)} & \textbf{12} & \textbf{(12 + 21)} & \textbf{19} & \textbf{(19 + 13)} & \textbf{28} & \textbf{(28 + 5)} & \textbf{70} \\

  \bottomrule
  \end{tabularx}
  
  \label{tab:RQ2}
\end{table*}

\subsubsection{Fine-Grained Effectiveness at the Function and Sequence Levels}
We decompose pre-screening into its two stages. At the function level, across the 160 function-level incidents, DeFiScreener recovers every labeled vulnerable function and thus attains 100\% recall with no omissions. All 38 false positives of the pipeline originate at this stage; as analyzed above, they stem from benign functions whose semantics closely resemble a vulnerable pattern.

Table \ref{tab:RQ1-2} presents the screening results for 47 real-world DeFi attack incidents involving vulnerable function call sequences. Across the 47 evaluated incidents, the dataset encompasses a total of 696 smart contracts and 4,121 functions, which constitute an expansive search space of 13,730 potential call sequences. Out of the 55 ground-truth vulnerable sequences, DeFiScreener effectively identified 51 TP, yielding a 92.73\% recall. Interestingly, during the sequence screening phase, DeFiScreener achieved zero complete misses (FN) across all 47 incidents.

It is also worth noting that no screened sequence in this evaluation is entirely unrelated to a ground-truth sequence, i.e., we observe no genuine sequence-level false positives. This partly reflects our evaluation protocol, which caps the number of screened sequences per incident at its number of ground-truth sequences. Within this budget, DeFiScreener's output still overlapped with a ground-truth sequence, fully (TP) or partially (MP), in every one of the 47 incidents.

The evaluation identified four instances of MPs, which represent incomplete fragments of vulnerable sequences. These omissions originated from four protocols: CloberDEX \cite{CloberDEX}, Cover Protocol \cite{CoverProtocol}, PancakeHunny \cite{PancakeHunny}, and TreasureDAO \cite{TreasureDAO}. For \textit{CloberDEX}, the \textit{lock} function is implemented externally via the \textit{IBookManager.sol} interface. Consequently, from a single-protocol perspective, the internal logic of the \textit{lock} function remains opaque, leading to a discontinuity in the \textit{burn-lock-lockAcquired} sequence and resulting in the capture of only the \textit{burn} function. Regarding \textit{TreasureDAO}, the presence of two \textit{buyItem} functions across different contracts caused DeFiScreener to incorrectly prioritize the proxy contract code while failing to extract the underlying core logic. The underlying factors for the omissions in \textit{Cover Protocol} and \textit{PancakeHunny} are multifaceted and are elaborated upon in Appendix \ref{appendix:RQ1}.

\subsection{RQ2: Pre-screening Feasibility}
RQ2 examines whether DeFiScreener generalizes to previously unseen attacks, rather than only retrieving incidents already stored in its library. To this end, we designed a controlled zero-day scenario. As shown in Table~\ref{tab:RQ3-1}, the left column lists historical incidents retained in the Attack Pattern Library. The right column lists later target protocols that were fully held out during evaluation. For each target, the incident itself and its entire fork were excluded from the library. Each success therefore demonstrates generalization to an unseen contract, since the matched pattern originated from a separate, earlier incident rather than the target.

\subsubsection{Construction of the evaluation pairs}
Each pair links a historical incident to a later protocol that reproduced the same attack pattern. We anchored this pairing in public post-mortem and audit reports from SlowMist~\cite{slowmist} and DeFiHackLabs~\cite{DeFiHackLabs}, which document the root cause of each incident. Two PhD researchers with smart-contract security expertise independently inspected every candidate pair. Each researcher labeled whether the two incidents shared the same underlying vulnerability, using the reports and the contract source as evidence. When the two researchers disagreed, they resolved the case by discussion, and a senior researcher adjudicated the few unresolved conflicts.

\subsubsection{Zero-Day Attack Screening Efficacy}
Across the 17 held-out pairs, DeFiScreener pinpointed the exact vulnerable function or sequence in 14 targets (\ding{51}). A single pattern often generalized to several unseen protocols. The \textit{deliver} pattern extracted from the TomInu Token~\cite{TomInuToken} incident recovered the vulnerability in BEVO~\cite{BEVO}, FDP Token~\cite{FDPToken}, and OLIFE~\cite{OLIFE}. Likewise, the \textit{balanceOf} pattern from NOVO Protocol~\cite{NOVO_Protocol} screened four later tokens, including ROI~\cite{ROI} and Sheep Token~\cite{Sheep_Token}. Because every target and its fork were held out, each hit reflects transfer from an earlier, structurally analogous incident.


Furthermore, DeFiScreener achieved a partial screening result ($\bullet$) in the PancakeHunny \cite{PancakeHunny}. While DeFiScreener recognized anomalous semantic features associated with the \textit{getReward} function (PM-1), the complete vulnerable function call sequence is not entirely isolated. The root cause is consistent with that detailed in Appendix \ref{appendix:RQ1}.

Two targets were missed (\ding{55}): CloberDEX~\cite{CloberDEX} and YIEDL~\cite{YIEDL}. Both misses stem from the target architecture, not from the matching algorithm. The \textit{lock} function in CloberDEX and the \textit{redeem} function in YIEDL are interface declarations without a function body, so no implementation semantics were available to match.

Despite these fail cases, the empirical results validate DeFiScreener's efficacy in leveraging historical data to uncover unreported, structurally analogous vulnerabilities prior to active exploitation.

\subsection{RQ3: Discovering Downstream Misses}
RQ3 asks whether historical-matching pre-screening adds coverage to specialized downstream detectors. In this setting, DeFiScreener runs as a pre-screen ahead of each detector, flagging candidate vulnerable functions and sequences by matching against the Attack Pattern Library. The pipeline reports an incident if either the detector or DeFiScreener flags it. We write $\uparrow$ for an incident that DeFiScreener recovers but the standalone detector misses. This recovery is possible because each specialized detector is bound to a single technique and attack class, and therefore has systematic blind spots. Historical matching, by contrast, flags any contract that reuses a disclosed pattern, independent of the downstream technique. Consistent with our deployment scenario, the pre-screen operates on patterns already disclosed and stored in the library. RQ3 therefore measures how many known-pattern attacks pre-screening recovers beyond what each downstream detector catches alone, whereas generalization to unseen contracts is evaluated separately, without this stored information, in RQ2. We conduct the evaluation on Dataset~\uppercase\expandafter{\romannumeral 2} (price manipulation) and Dataset~\uppercase\expandafter{\romannumeral 3} (logic vulnerabilities).

\subsubsection{Effectiveness of Pre-screening on Price Manipulation Detection}
Table \ref{tab:RQ2} details the detection performance of four baseline tools (DeFiRanger \cite{wu2023defiranger}, DeFiTainter \cite{kong2023defitainter}, DeFort \cite{xie2024defort}, and DeFiScope \cite{zhong2025detecting}) when operating independently and when combined with DeFiScreener for pre-screening. We observe that existing baseline tools exhibit varying degrees of false negatives in standalone operation; specifically, DeFiRanger, DeFiTainter, and DeFort correctly identified only 14, 12, and 19 attack incidents, respectively.

However, pre-screening with DeFiScreener discovers a substantial number of incidents that each downstream tool misses on its own. DeFiScreener independently flags 19, 21, 13, and 5 additional true incidents ($\uparrow$) for DeFiRanger, DeFiTainter, DeFort, and DeFiScope, respectively, none of which the corresponding tool detects when run alone. Across Dataset~\uppercase\expandafter{\romannumeral 2}, the pre-screening mechanism recovers a total of 70 incidents that would have been completely overlooked had we relied solely on these state-of-the-art analyzers. Notably, for tools with lower standalone detection rates, such as DeFiTainter, pre-screening substantially expands the combined coverage, raising the total number of identified incidents from 12 to 33. Even for the best-performing baseline tool, DeFiScope, pre-screening recovers 5 additional incidents that it alone misses.

We observed that pre-screening discovered no additional incidents in two cases: PancakeHunny \cite{PancakeHunny} and APEDAO \cite{APEDAO}. For PancakeHunny, this null result matches the limitation identified in RQ1, where DeFiScreener isolates only partial fragments of the vulnerable sequence (see Appendix \ref{appendix:RQ1}). For APEDAO, the miss stems from source-code unavailability, as noted in the dataset description.

\subsubsection{Effectiveness of Pre-screening on Logic Vulnerability Detection}
We evaluate how effectively DeFiScreener recovers logic vulnerabilities missed by downstream detectors on Dataset \uppercase\expandafter{\romannumeral 3}. Because the conclusions match those for price manipulation detection, we detail this experiment in Appendix \ref{appendix:RQ2}.

\subsection{RQ4: Ablation Study}
To evaluate the contributions of DeFiScreener's core components and assess the hyperparameter sensitivity, we performed a comprehensive ablation study.

\begin{figure}[!t]
  \includegraphics[width=3.5 in]{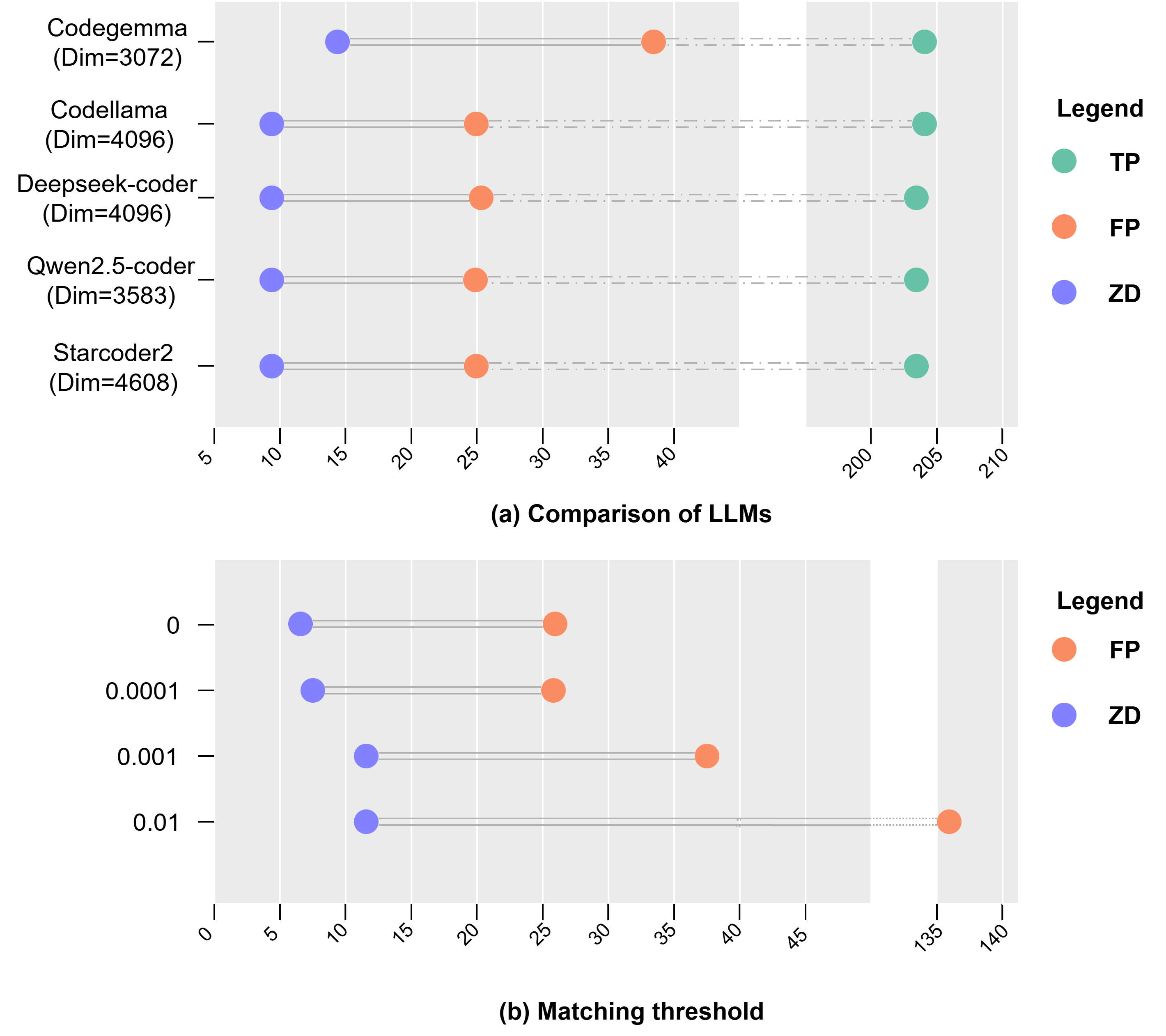}
  \caption{Ablation studies on different LLMs (a) and matching thresholds (b). The LLM ablation is comprehensively evaluated at both the function and sequence levels, whereas the matching threshold ablation is evaluated at the function level. ZD is the number of held-out zero-day targets from RQ2 recovered under each configuration.}
  \label{fig:RQ4-1}
\end{figure}

\subsubsection{Robustness of Semantic Modeling}
We ablate the semantic-extraction stage over five code-specialized LLMs. In Fig.~\ref{fig:RQ4-1}(a), ZD is the number of held-out zero-day targets from RQ2 (out of 17), a measure of generalization rather than in-library retrieval. The five backbones behave consistently. Each attains a true-positive count of at least 203 out of 207 and recovers at least nine held-out targets, so DeFiScreener does not hinge on a specific encoder. We adopt Codegemma as the primary encoder for three reasons. First, it attains the highest true-positive count (204), sacrificing no recall. Second, it has the smallest embedding dimension (3072) and therefore the lowest matching and memory cost, in keeping with our efficiency goal. Third, because DeFiScreener is a pre-screen whose candidates are verified by downstream detectors, its comparatively higher false-positive count is inexpensive, so we prioritize recall and efficiency over standalone precision. Consistent with these choices, Codegemma also yields the strongest held-out generalization (ZD = 14 versus 9).

\begin{figure*}[!t]
  \includegraphics[width=7 in]{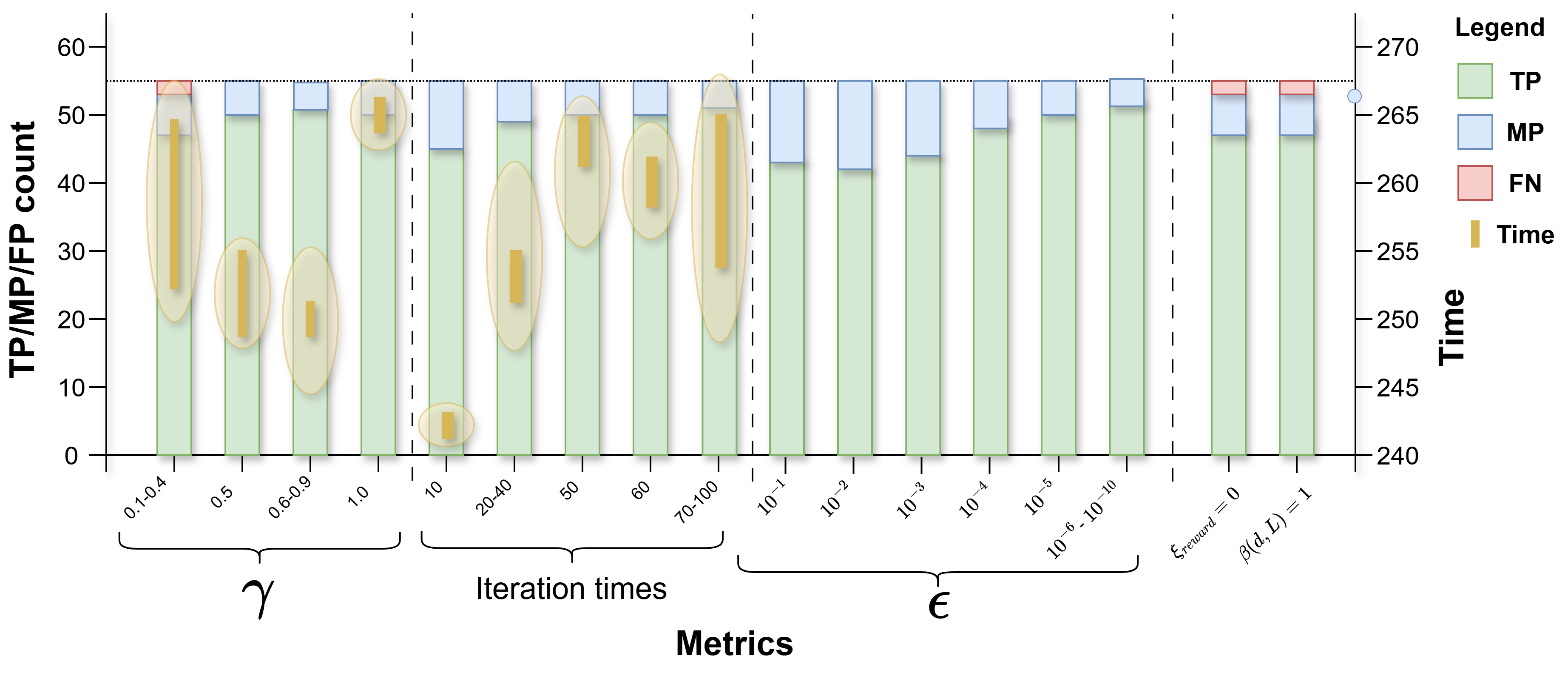}
  \caption{Parameter sensitivity analysis of APO-MCTS. The total number of vulnerable sequences is 55 (indicated by the top dashed line). MP is the number of partially captured vulnerable sequences. The ellipse plots and their internal vertical lines represent the coverage range and high-frequency regions of the search time, respectively.}
  \label{fig:RQ4-2}
\end{figure*}

\subsubsection{Function-Level Matching Threshold}
The matching threshold controls the precision of vulnerable-function matching, and we set it to 0.001. Across all four threshold values tested, DeFiScreener matches all 160 function-level ground-truth vulnerable functions. Fig.~\ref{fig:RQ4-1}(b) therefore reports only ZD and FP, the two metrics that actually vary with the threshold. Raising it from 0.001 to 0.01 does not improve generalization, since ZD stays at 12, but inflates false positives from 38 to 136. Tightening it below 0.001 instead lowers ZD, from 12 to 8 at 0.0001. The value 0.001 therefore gives the best precision at the point where generalization saturates.

\subsubsection{APO-MCTS Ablation}
\textbf{Hyperparameter Sensitivity.} Fig.~\ref{fig:RQ4-2} reports TP, MP, and FP for $\gamma$, $\epsilon$, and the iteration budget, out of 55 ground-truth vulnerable sequences. An excessively small discount factor $\gamma$ (0.1-0.4) compels the APO-MCTS algorithm to frequently oscillate its search space across various root nodes, lowering TP to 47 and producing 2 complete misses (FN). Configuring $\gamma = 1$ instead rigidly confines the exploration to a single root node, capping TP at 50. Furthermore, an excessively large $\epsilon$ diminishes the numerical distinguishability between closely related semantics, thereby leading to the excessive extraction of partial vulnerable call sequences. MP peaks at 12 of 55 when $\epsilon = 10^{-1}$ and falls to 4 once $\epsilon$ tightens to between $10^{-6}$ and $10^{-10}$. Finally, when the number of search iterations is insufficient, the algorithm prematurely executes the optimal sequence selection before complex \textsf{FCT}s are fully expanded, which compromises the detection of complete vulnerable sequences.

Fig.~\ref{fig:RQ4-2} also shows that search time falls for $\gamma$ as the value rises toward $0.6$-$0.9$, reflecting less oscillation across root nodes, but climbs again at $\gamma = 1$ once the search is confined to a single root node. For the iteration budget, time instead rises quickly at first, then increases only marginally thereafter.

Consequently, DeFiScreener achieves optimal convergence when the $\gamma$ is configured within the range of $0.6$ to $0.9$, the $\epsilon$ is set strictly between $10^{-6}$ and $10^{-10}$, and the search undergoes $70$ to $100$ iterations.

\textbf{Algorithm Components. } When $\xi_{reward} = 0$ or deactivating the penalty-based score decay mechanism ($\beta(d, L) = 1$) substantially degrades search efficacy, reducing the TP count to 47 while producing complete misses (FN). This empirically substantiates that standard MCTS is inherently more susceptible to overlooking critical vulnerable function call sequences.


\section{DISCUSSION}
\subsection{Efficient Pre-screening of Vulnerable Instances}
Serving as an efficient pre-screening tool, DeFiScreener effectively triages potential DeFi attack vectors within smart contracts and directs specialized downstream detectors toward targeted analysis, thereby significantly fortifying the smart contract security ecosystem. Furthermore, by leveraging its historical attack pattern matching capabilities, DeFiScreener enables the agile scanning of unpatched legacy contracts following the public disclosure of novel vulnerabilities, preemptively mitigating threats before adversaries can weaponize these flaws.

\subsection{Limitations of DeFiScreener}
\label{subsec:limitations}
While DeFiScreener effectively screens vulnerable instances within DeFi protocols, our current study presents certain limitations that concurrently outline directions for future research.

\textbf{Static, Single-Protocol Analysis.} DeFiScreener reasons over a static snapshot of a single protocol's source code, which introduces two related boundaries. First, it does not model the dynamic runtime state or transaction context, so it can over-approximate the exploitability of state-dependent operations. Second, it does not follow calls whose implementation lies in another protocol or is unavailable as source, so interface-only functions and cross-protocol interactions fall outside its scope. Future work will incorporate transaction-level behavioral features into the embeddings and integrate external protocol sources to relax both boundaries.

\textbf{Potential Omission of Novel Attack Patterns.} As DeFiScreener's screening mechanism relies on an attack pattern library constructed from historical data, it may omit completely novel attack vectors. Nevertheless, a substantial corpus of empirical attack cases demonstrates a high recurrence rate of attack patterns within the DeFi ecosystem, ensuring that DeFiScreener retains significant practical utility.

\section{RELATED WORK}
\subsection{General Smart Contract Vulnerability Detection}
Securing smart contracts against malicious intrusions necessitates rigorous security auditing prior to deployment. To this end, the research community has proposed various general smart contract vulnerability detection tools designed to uncover multiple vulnerabilities within the contracts.

\textit{Slither} \cite{feist2019slither} translates smart contracts into a specialized intermediate representation to facilitate automated vulnerability detection using techniques such as data-flow and taint tracking.
\textit{Securify2} \cite{securify2} compiles source code into a customized intermediate representation and performs context-sensitive pattern matching for vulnerability detection.
\textit{ConFuzzius} \cite{torres2021confuzzius} synergistically integrates evolutionary fuzzing with constraint solving to navigate complex execution paths, and leverages dynamic data dependency analysis to efficiently construct multi-transaction sequences for deep vulnerability detection.
\textit{sFuzz} \cite{nguyen2020sfuzz} introduces an efficient and lightweight multi-objective adaptive strategy to uncover potential vulnerabilities hidden in hard-to-cover branches. 
\textit{Smartian} \cite{choi2021smartian} leverages static analysis to extract semantic dependencies for generating initial transaction sequences, and incorporates lightweight dynamic data-flow feedback to systematically explore deep contract states for vulnerability detection.
\textit{TMP} \cite{zhuang2021smart} builds a temporal message propagation graph model by transforming the source code of smart contracts into a normalized contract graph, thus enabling the detection of contract vulnerabilities.

\subsection{Specialized DeFi Attack Detection}
While general detection methods identify conventional vulnerabilities, they lack the semantic awareness required for complex DeFi logic. In recent years, researchers have developed dedicated tools targeting specific exploits, such as price manipulation and logic flaw.
\subsubsection{Price Manipulation Attack}
\textit{DeFiRanger} \cite{wu2023defiranger} reconstructs the high-level semantics of DeFi protocols by building cash flow trees from transactions and elevating low-level semantics to a higher level. It then uses the recovered semantic representation to detect price manipulation attacks.
\textit{DeFiTainter} \cite{kong2023defitainter} constructs tainted-call graphs among contracts based on code constants and function parameters, and it deduces the high-level semantics of price manipulation attacks. 
\textit{DeFort} \cite{xie2024defort} establishes a general behavior model to systematically monitor abnormal token price fluctuations and leverages multi-strategy profit calculation mechanisms to automatically trace price manipulation attacks in DeFi applications.
\textit{DeFiScope} \cite{zhong2025detecting} leverages a fine-tuned LLM to abstract custom price calculation logic from smart contracts, and integrates recovered high-level on-chain operations with systematically mined behavior patterns to detect diverse price manipulation attacks.

\subsubsection{Business Logic Flaw}
\textit{GPTScan} \cite{sun2024gptscan} decomposes logic vulnerabilities into predefined scenarios and properties, leveraging GPT for semantic-aware candidate matching followed by rigorous static confirmation to detect smart contract defects.
\textit{FORAY} \cite{wen2024foray} leverages a domain-specific language to abstract smart contracts into high-level financial operations, and models DeFi protocols as token flow graphs to systematically synthesize deep logical exploits via targeted reachability analysis.


These methods have advanced DeFi security, yet each category has a structural limitation. General analyzers (e.g., Slither, ConFuzzius) target instruction-level bugs and lack the semantic depth to capture protocol-level DeFi logic. Specialized detectors (e.g., DeFiRanger, DeFort, DeFiScope, FORAY) are precise, but each covers a single attack class, so comprehensive coverage requires assembling and maintaining many fragmented tools. DeFiScreener does not compete with these detectors; it addresses their coverage gap as a general pre-screen. By matching against historical exploits, it flags candidates across attack classes, independent of any single detection technique, and routes them to the appropriate detector. RQ3 confirms this empirically: pre-screening recovers a total of 88 attacks across 7 state-of-the-art detectors.


\section{CONCLUSION}
In this paper, we propose DeFiScreener, the first automated pre-screening framework designed to address the coverage limitations of existing DeFi security tools by leveraging historical attack cases. DeFiScreener leverages an LLM to extract semantic embeddings and constructs \textsf{FCT}s to facilitate a dual-level screening mechanism. Through a large-scale empirical evaluation on a dataset of 207 real-world attack incidents representing approximately \$2.97 billion in total losses, DeFiScreener demonstrated remarkable effectiveness in DeFi attack pre-screening. This effectiveness highlights its practical value that discovers attacks otherwise missed by downstream detectors. By complementing rather than replacing those detectors, DeFiScreener helps proactively secure the rapidly evolving DeFi ecosystem.

\bibliographystyle{IEEEtran}
\bibliography{ref}

\appendices

\begin{figure}[htbp]
  \centering
  \begin{lstlisting}
// Shared deliver function (TomInu Token / BEVO / FDP Token / OLIFE)
function deliver(uint256 tAmount) public {
    address sender = _msgSender();
    require(!_isExcluded[sender], "Excluded addresses             cannot call this function");
    (uint256 rAmount,,,,,,) = _getValues(tAmount);
    _rOwned[sender] = _rOwned[sender].sub(rAmount);
    _rTotal = _rTotal.sub(rAmount);   // shrinks the rate
    _tFeeTotal = _tFeeTotal.add(tAmount);
}
  \end{lstlisting}
  \caption{Shared \texttt{deliver} function inherited from the Reflect Finance template across TomInu Token, BEVO, FDP Token, and OLIFE.}
  \label{fig:deliver}
\end{figure}

\begin{figure*}[!t]
  \includegraphics[width=7 in]{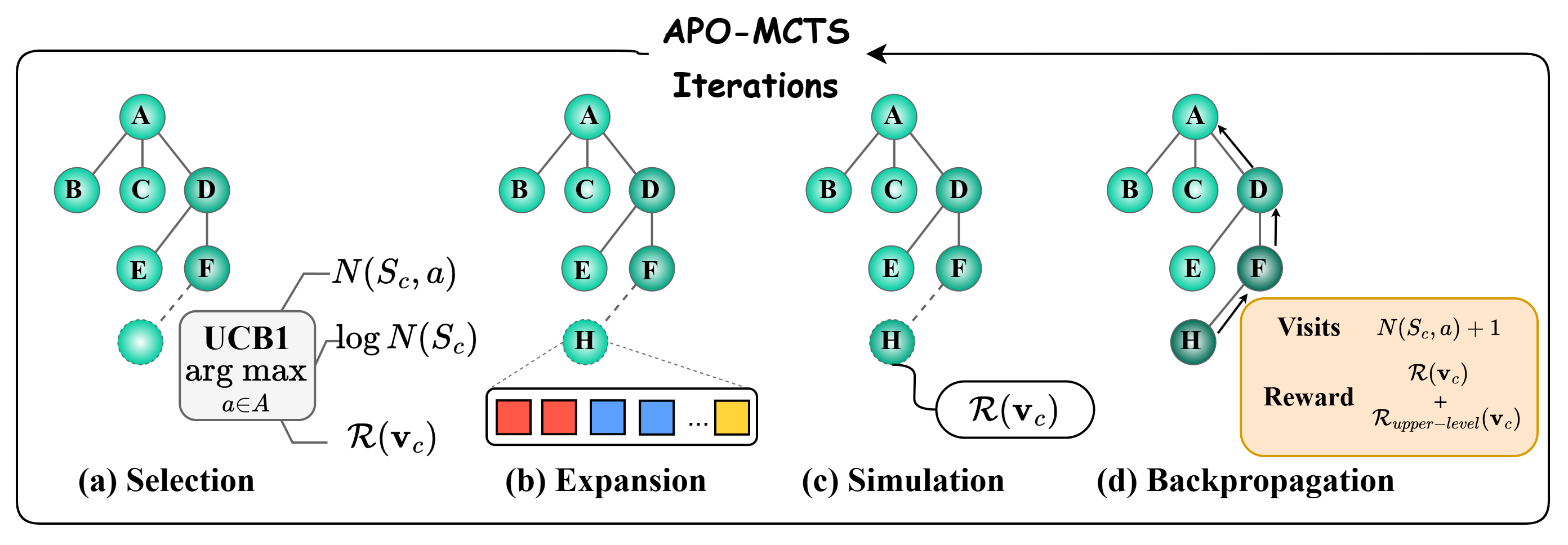}
    \caption{\centering APO-MCTS Iteration.}
  \label{fig:APO-MCTS Iteration}
\end{figure*}

\section{Case Study II: Serial Template Exploitation}
\label{appendix:motivation}
Another sobering illustration of exploit recurrence is the series of attacks targeting the \textit{reflection token} mechanism in \textit{TomInu Token} \cite{TomInuToken}, \textit{BEVO} \cite{BEVO}, \textit{FDP Token} \cite{FDPToken}, and \textit{OLIFE} \cite{OLIFE}. In this design, each holder's balance is not stored directly; instead, it is computed on-the-fly as $b = r_{\mathit{owned}} / \mathit{rate}$, where $\mathit{rate} = r_{\mathit{total}} / t_{\mathit{total}}$. Throughout 2023, these four independent BNB Smart Chain projects were each exploited via an attack rooted in this shared vulnerability.

As shown in Fig.~\ref{fig:deliver}, \texttt{deliver} decrements \texttt{\_rTotal} (line~8) while leaving \texttt{\_tTotal} unchanged (global variable), which shrinks the rate and causes every non-excluded holder's reflected balance to inflate without any actual token transfer. An attacker who first acquires a large token position via a flash loan can therefore call \texttt{deliver} to manufacture a balance discrepancy in the pair's computed balance, then call the standard AMM function to withdraw the discrepancy as real tokens. In the TomInu Token incident, this sequence drained approximately \$51,000 in liquidity, a pattern reproduced across BEVO, FDP Token, and OLIFE.

All four affected tokens descended from the same Reflect Finance contract template~\cite{reflectfinance} distributed through DxSale and similar launchpad services, and none independently audited the security interaction between \texttt{deliver} and the AMM pair's balance accounting. The three-month span from TomInu to OLIFE, during which the identical attack was re-executed against three derivative protocols, underscores that pattern-based vulnerability propagation is not an isolated coincidence but a structural property of the DeFi development ecosystem.

\section{APO-MCTS Iterations}
\label{appendix:APO-MCTS}
APO-MCTS iterations include selection, expansion, simulation and backpropagation, as shown in Fig.~\ref{fig:APO-MCTS Iteration}.

\textbf{State Space.} The state space $S_c$ is represented by \textsf{FCT} structure, which utilizes tree state to track historical trajectories and provide useful information for future decisions. 

\textbf{Action Space.} The actions $a$ are employed to modify the current architecture of \textsf{FCT}, dynamically adapting it based on the current state.

\textbf{Selection.} Starting from the root node, we traverse the current \textsf{FCT} structure. Upon reaching a leaf node, if it has unexpanded child nodes, the algorithm proceeds to the expansion step and adds new leaf nodes. The selection operations are performed using the UCB1 algorithm \cite{auer2002finite}:
 
\begin{equation}
UCB1(S_c) = \underset{a \in \textit{A}}{\text{arg max}}(\frac{\mathcal{R}(\mathbf{v}_c)}{N(S_c, a)} + \textit{C} \sqrt{\frac{2 * \log N(S_c)}{N(S_c, a)}})
\end{equation}
where $A$ indicates the set of all possible actions, $N(S_c, a)$ represents the number of times the contract node takes action $a$, and $N(S_c)$ is the number of visits to the state $S_c$ (i.e. $N(S_c) = \Sigma_{a \in A} N(S_c, a)$), and \textit{C} is a hyperparameter which tunes approximate accuracy and explores rare nodes. 

\textbf{Expansion.} In the expansion step, we select all possible actions to create new nodes, add the resulting states as child nodes in the tree, and initialize their statistical information. If an expanded node has no available actions $\dot{a}$, it is considered a terminal state $\dot{S_c}$, and the algorithm moves to the backpropagation step.

\textbf{Simulation.} We iteratively perform selection and expansion, sampling descendants of newly expanded nodes. To prevent infinite loops caused by cycles in function calls, we maintain a list of explored paths. If an ancestor node is encountered, the algorithm skips to the next iteration.

\textbf{Backpropagation.} In the backpropagation phase, the process begins at a leaf node and traverses back to the root node, updating visitation statistics along the way. Information is iteratively propagated to all ancestor nodes (i.e., invoking function nodes) until the root is reached.


\begin{figure}[!t]
  \centering
  \includegraphics[width=2.5 in]{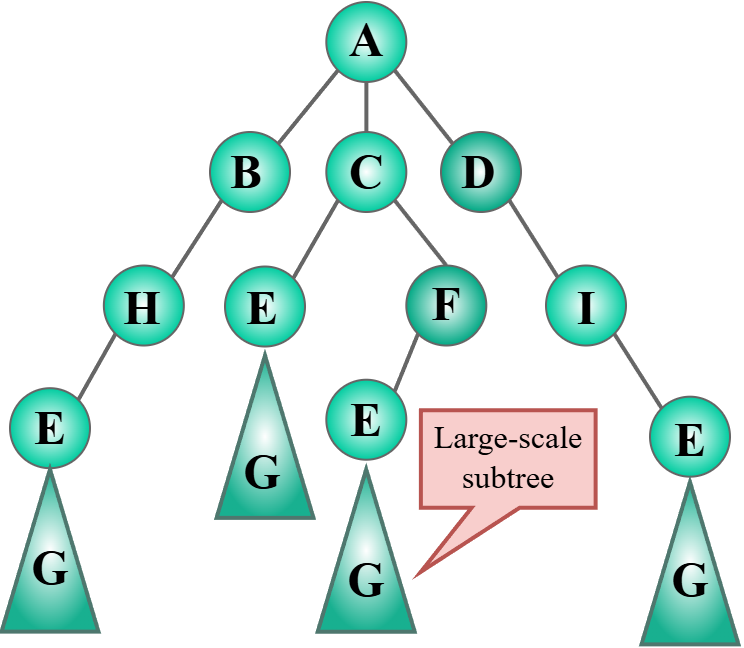}
  \caption{Case Study on Omitted Sequences: PancakeHunny.}
  \label{fig:PancakeHunny}
\end{figure}

\begin{table*}[t]
  \centering
  
  \caption{Comparison of results between running downstream tools independently and incorporating pre-screening, in logic vulnerability.}

  \begin{tabularx}{\textwidth}{c l l c c c c c c c c c }
  \toprule 
  \textbf{\#} & \textbf{Protocol} & \textbf{Date} & \textbf{Loss} & \textbf{Chain} & \textbf{Foray} & \textbf{DSc+FORAY} & \textbf{ItyFuzz} & \textbf{DSc+ItyFuzz} & \textbf{Halmos} & \textbf{DSc+Halmos} & \textbf{UP($\uparrow$)} \\
  \midrule

1 & NOVO Protocol & 2022/05/29 & 76K & BSC & \ding{51} &  \ding{51} & \ding{51} & \ding{51} & \ding{55} & $\uparrow$ & \textbf{1} \\

2 & Discover & 2022/06/06 & 11K & BSC & \ding{51} & \ding{51} & \ding{55} & $\uparrow$ & \ding{51} & \ding{51} & \textbf{1} \\

3 & EGD Finance & 2022/08/07 & 36K & BSC & \ding{51} & \ding{51} & \ding{51} & \ding{51} & \ding{55} & $\uparrow$ & \textbf{1} \\

4 & NXUSD & 2022/09/06 & 50K & BSC & \ding{55} & $\uparrow$ & \ding{55} & $\uparrow$ & \ding{55} & $\uparrow$ & \textbf{3} \\

5 & BXH & 2022/09/28 & 40K & BSC & \ding{51} & \ding{51} & \ding{55} & $\uparrow$ & \ding{55} & $\uparrow$ & \textbf{2} \\

6 & AES & 2022/12/07 & 60K & BSC & \ding{51} & \ding{51} & \ding{51} & \ding{51} & \ding{55} & $\uparrow$ & \textbf{1} \\

7 & MU\&MUG & 2022/12/10 & 57K & AVAX & \ding{51} & \ding{51} & \ding{55} & $\uparrow$ & \ding{51} & \ding{51} & \textbf{1} \\
  
8 & Lodestar & 2022/12/11 & 4M & ARB & \ding{55} & $\uparrow$ & \ding{55} & $\uparrow$ & \ding{55} & $\uparrow$ & \textbf{3} \\
  
9 & BGLD & 2022/12/12 & 18K & BSC & \ding{51} & \ding{51} & \ding{51} & \ding{51} & \ding{55} & $\uparrow$ & \textbf{1} \\
  
10 & Axioma & 2023/04/24 & 13K & BSC & \ding{51} & \ding{51} & \ding{55} & $\uparrow$ & \ding{55} & $\uparrow$ & \textbf{2} \\
  
11 & SellToken02 & 2023/05/13 & 197K & BSC & \ding{51} & \ding{51} & \ding{55} & $\uparrow$ & \ding{55} & $\uparrow$ & \textbf{2} \\
  \midrule
  
  \# & Total (\ding{51} + $\uparrow$) & \textbf{-} & \textbf{-} & \textbf{-} & \textbf{9} & \textbf{(9 + 2)} & \textbf{4} & \textbf{(4 + 7)} & \textbf{2} & \textbf{(2 + 9)} & \textbf{18} \\

  \bottomrule
  \end{tabularx}
  
  \label{tab:RQ2-2}
\end{table*}

\section{Analysis of Partial Sequence Screening in Cover Protocol and PancakeHunny}
\label{appendix:RQ1}
\subsection{Analysis of PancakeHunny}
As illustrated in Fig.~\ref{fig:PancakeHunny}, the \textsf{FCT} rooted at node A exhibits significant path explosion and state-space expansion. Specifically, function E is triggered with high frequency across multiple distinct invocation sequences (e.g., acting as a subsequent call for nodes H, C, F, and I). It is particularly noteworthy that every invocation of function E inevitably triggers a large-scale subtree rooted at G.

During the expansion phase of the APO-MCTS algorithm utilized by PancakeHunny, this specific topological structure introduces a severe computational bottleneck. Given the strict upper bound on the number of iterations in APO-MCTS, the recurrent exploration of the massive subtree G rapidly and heavily consumes the limited search budget. This not only prevents the complete traversal and expansion of the entire \textsf{FCT} rooted at A within a finite iteration cycle but also leads to a severe skew and imbalance in the allocation of search resources. Consequently, sibling branches ordered later in the sequence (such as function D, a direct child of root node A) may suffer from starvation, failing to even commence expansion before the iteration cycle terminates.

\begin{figure}[!t]
  \includegraphics[width=3.5 in]{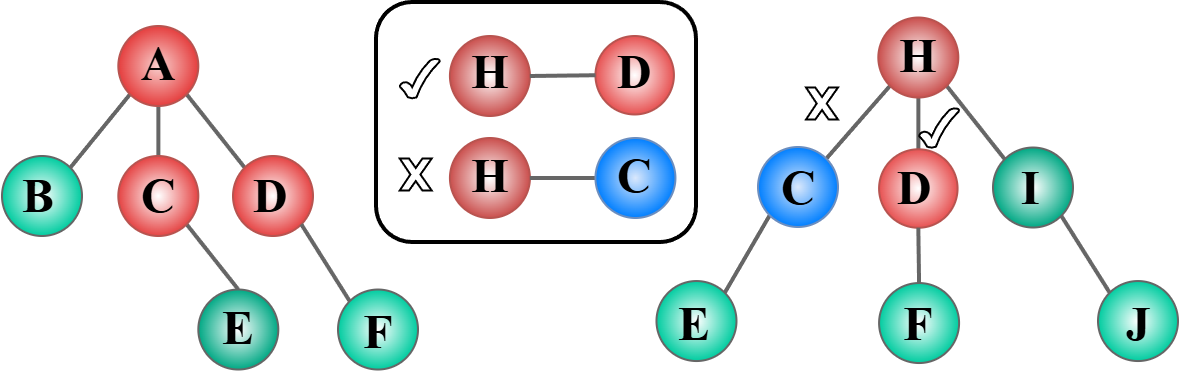}
  \caption{Case Study on Omitted Sequences: Cover Protocol.}
  \label{fig:Cover Protocol}
\end{figure}

Because the redundant computation of these massive subtrees prematurely depletes the search budget, PancakeHunny is forced to advance to the screening of vulnerable function call sequences before the \textsf{FCT} rooted at A is fully expanded. This ultimately restricts DeFiScreener to capturing and screening only partial fragments of vulnerable sequences, thereby failing to guarantee the completeness of the vulnerable sequence screening.

To mitigate the computational bottleneck induced by redundant subtree expansions, our future work will explore the integration of a state-sharing mechanism within the APO-MCTS. Specifically, recurrently invoked large-scale subtrees (such as subtree $G$) will be modeled as cross-path shared structural components. Upon its initial comprehensive expansion, the internal topological state and associated heuristic search metrics of the large-scale subtree will be globally cached by APO-MCTS. During subsequent invocations of the same subtree across different branch paths, the algorithm will directly retrieve and reuse its pre-computed cached state, thereby bypassing redundant exploration processes. This optimization strategy is designed to significantly conserve the limited MCTS search budget, facilitate the complete traversal of the global \textsf{FCT}.

\subsection{Analysis of Cover Protocol}
In the evaluation of the Cover Protocol, DeFiScreener efficiently identified partial vulnerable segments but exhibited a specific deviation during the final sequence extraction phase. According to the ground truth, the protocol contains exactly three vulnerable sequences: A-C, A-D, and H-D (denoted by the red nodes in Fig.~\ref{fig:Cover Protocol}). However, the APO-MCTS algorithm output A-C, A-D, and H-C. While the first two sequences were correctly identified, the true sequence H-D was omitted in favor of H-C.

An in-depth analysis of the \textsf{FCT}s for the Cover Protocol reveals that this deviation stems from a heuristic status-carryover issue during the node evaluation process. In the \textsf{FCT} rooted at A, node C is correctly identified as a vulnerable function, prompting the APO-MCTS to assign it a high vulnerability confidence score. Consequently, when the algorithm subsequently explores the \textsf{FCT} rooted at H, the historical vulnerability weighting of node C skews the selection phase, causing the search to greedily prioritize the H-C path.

Furthermore, the experimental design for RQ1 enforces a strict cardinality constraint on the output: the maximum number of screened sequences is strictly limited to the exact number of ground-truth vulnerabilities. Because the algorithm prematurely converges on H-C and fulfills its strict quota of three sequences, the search process is truncated. This ultimately causes the true vulnerable node D in the second tree to be missed.

In future research, we aim to refine the APO-MCTS by incorporating a holistic path-matching mechanism. Specifically, during the exploration of child nodes, the algorithm will dynamically assess whether the entire sequence connected to the parent node strictly aligns with known vulnerable sequences within the attack pattern library, thereby preventing anomalous screening such as the H-C sequence.

\section{Effectiveness of Pre-screening on Logic Vulnerability
Detection}
\label{appendix:RQ2}
Table \ref{tab:RQ2-2} details the detection performance of three baseline tools (Foray \cite{wen2024foray}, ItyFuzz \cite{shou2023ityfuzz}, and Halmos \cite{Halmos}) on logic vulnerabilities when running independently and when combined with DeFiScreener for pre-screening. We observe that existing baseline tools exhibit significant false negatives in standalone operation; specifically, Foray, ItyFuzz, and Halmos correctly identified only 9, 4, and 2 attack incidents out of the 11 evaluated cases, respectively.

However, pre-screening with DeFiScreener recovers a substantial number of incidents that each downstream tool misses on its own. DeFiScreener introduces 2, 7, and 9 new valid alerts ($\uparrow$) for Foray, ItyFuzz, and Halmos, respectively, none of which the corresponding tool detects when run alone. Across Dataset \uppercase\expandafter{\romannumeral 3}, the pre-screening mechanism generates a total of 18 critical warnings for vulnerabilities that would have been completely overlooked had we relied solely on the standalone analyzers. Notably, for tools with lower initial detection rates (Halmos and ItyFuzz), pre-screening substantially expands their effective coverage, increasing the total number of identified incidents from 2 to 11 and from 4 to 11, respectively. Furthermore, even for the best-performing baseline tool (Foray), the pre-screening mechanism effectively captures 2 additional, highly concealed attack incidents.

\end{document}